\documentclass[preprint2]{aastex61}

\submitjournal{ApJ}

\shorttitle{Ly$\alpha$ Emitters at $z\sim2.23$}
\shortauthors{Hao et al.}

\begin{document}

\title{A Deep Ly$\alpha$ Survey in ECDF-S and COSMOS: I. General Properties of Ly$\alpha$ Emitters at $z\sim2$}

\correspondingauthor{Cai-Na Hao}
\email{hcn@bao.ac.cn}

\author{Cai-Na Hao} 
\affiliation{Tianjin Astrophysics Center, Tianjin Normal University, Tianjin 300387, China}

\author{Jia-Sheng Huang}
\affiliation{National Astronomical Observatories, Chinese Academy of Sciences, Beijing 100012, China}
\affiliation{Harvard-Smithsonian Center for Astrophysics, 60 Garden Street, MS65, Cambridge, MA 02138, USA}

\author{Xiaoyang Xia}
\affiliation{Tianjin Astrophysics Center, Tianjin Normal University, Tianjin 300387, China}

\author{Xianzhong Zheng}
\affiliation{Purple Mountain Observatory, Chinese Academy of Sciences, Nanjing 210008, China}

\author{Chunyan Jiang}
\affiliation{Key Laboratory for Research in Galaxies and Cosmology of Chinese Academy of Sciences, Shanghai Astronomical Observatory, Shanghai 200030, China}

\author{Cheng Li}
\affiliation{Physics Department and Tsinghua Center for Astrophysics, Tsinghua University, Beijing 100084, China}


\begin{abstract}

 Ly$\alpha$ Emitters (LAEs) may represent an important galaxy population in
the low mass regime. We present our deep narrowband imaging surveys in the
COSMOS and ECDF-S fields and study the properties of LAEs at $z=2.23\pm0.03$.
The narrowband surveys conducted at Magellan II telescope allow us to obtain a
sample of 452 LAEs reaching a $5\sigma$ limiting magnitude of $\sim26$ mag.
Our Ly$\alpha$ luminosity functions extend to $10^{41.8}$\,erg\,s$^{-1}$ with
steep faint-end slope. Using multi-wavelength ancillary data, especially
the deep \textit{Spitzer}/IRAC 3.6\,$\mu$m and 4.5\,$\mu$m photometric data, we
obtained reliable stellar mass estimates for 130 IRAC-detected LAEs, spanning a
range of $8 < {\rm log}(M_\star/M_\odot)< 11.5$.  For the remaining
IRAC-undetected LAEs, the median-stacked spectral energy distribution yields a
stellar mass of ${\rm log}(M_\star/M_\odot)=7.97^{+0.05}_{-0.07}$ and the
rest-frame ultraviolet emission indicates a median star formation rate of ${\rm
log} (SFR/M_\odot$\,yr$^{-1})=-0.14\pm0.35$. There are six LAEs detected by
the \textit{Spitzer}/MIPS 24\,$\mu$m or even \textit{Herschel} far-infrared
observations. Taking into account the six MIR/FIR detected LAEs, our LAEs cover
a wide range in the star formation rate (${\rm
1<SFR<2000}$\,M$_\odot$\,yr$^{-1}$). Although LAEs as a population are diverse
in their stellar properties, they are mostly low-mass star-forming
galaxies and follow the star formation main sequence relations or their
extrapolations to the low-mass end, implying a normal star-forming nature of
LAEs. The clustering analysis indicates that our LAEs reside in dark matter
halos with ${\rm <\log(M_{h}/M_{\odot})> =10.8^{+0.56}_{-1.1}}$, 
suggesting that they are progenitors of local Large Magellanic Cloud-like
galaxies. 

\end{abstract}





\section{INTRODUCTION}

The epoch at $z\sim2$ is crucial in the history of galaxy evolution when the
cosmic star formation rate density (SFRD) reaches its peak \citep[][and
references therein]{Madau+2014}.  Detailed knowledge of massive ($>10^{10}
M_{\odot}$) galaxies at this epoch has been widely investigated
\citep[e.g.,][]{Erb+2006,Forster+2009,Steidel+2014,Kriek+2015,Burkert+2016,Wuyts+2016}.
On the other hand, low-mass galaxies at $z\sim 2$ pose an unique and important
position in studying galaxy evolution because they are building blocks of local
mature galaxies.  However, our knowledge on low-mass galaxies ($<10^{10}
M_{\odot}$) at $z\sim2$ is still limited due to challenges in identifying these
faint galaxies.  Large samples, on the other hand, are needed for a robust
census of such galaxies.  The narrowband imaging technique is an effective way
of detecting Ly$\alpha$ emitting galaxies (LAEs) at specific redshifts
\citep[e.g.,][]{Malhotra+2002,Wang+2005,Finkelstein+2007,Finkelstein+2008,Gawiser+2007,
Gronwall+2007, Nilsson+2007,Pirzkal+2007, Lai+2008,
Ono+2010a,Ono+2010b,Acquaviva+2011, Zheng+2016}. Other methods such as
integral-field spectroscopy \citep[e.g.,][]{vanBreukelen+2005,Drake+2017}, slit
spectroscopy \citep[e.g.,][]{Rauch+2008} and medium-band imaging \citep[e.g.,][]{Stiavelli+2001,Taniguchi+2015,Sobral+2018} have also been employed in finding
LAEs. LAEs were found to be mostly composed of low-mass star-forming galaxies
\citep[e.g.,][]{Gawiser+2006, Finkelstein+2007, Lai+2008, Pirzkal+2007,
Nilsson+2011, Guaita+2011, Shimakawa+2017}, red massive star-forming LAEs exist
though \citep[e.g.,][]{Stiavelli+2001,Lai+2008,Finkelstein+2008,Finkelstein+2009,
Ono+2010b,Acquaviva+2011,Guaita+2011,Nilsson+2011,Oteo+2012,Matthee+2016}.  Therefore, LAEs can be used to probe
the properties of low-mass galaxies at high redshifts.

In the past several years, a number of narrowband imaging surveys and
spectroscopic observations have been carried out to search for LAEs at $z\sim2$
\citep[e.g.,][]{Nilsson+2009, Guaita+2010, Nakajima+2012, Blanc+2011,
Cassata+2011, Hathi+2016, Matthee+2016, Shimakawa+2017}. These surveys and
deep multi-wavelength ancillary data have made it possible
to yield measurements of properties of LAEs such as Ly$\alpha$ luminosity
function, star formation rate, stellar mass, dark matter halo mass and
rest-frame optical spectroscopic properties etc \citep[e.g.,][]{Guaita+2011,
Nilsson+2011,Ciardullo+2012, Ciardullo+2014,Guaita+2013,
Nakajima+2013, Trainor+2016, Sobral+2017,Kusakabe+2018}.

The Ly$\alpha$ luminosity function and its faint-end slope are of special
interest since they can serve as probes of galaxy evolution and cosmic
re-ionization \citep[e.g.,][]{Rauch+2008, Konno+2016, Zheng+2017}. In order to
determine the faint-end slope of the Ly$\alpha$ luminosity function, many
surveys have been carried out to detect LAEs with much faint luminosities
\citep[e.g.,][]{Blanc+2011, Ciardullo+2012}.  Based on a deep spectroscopic
survey, \citet{Cassata+2011} put strong constraints on the faint-end slope of
the Ly$\alpha$ luminosity functions at $1.95 < z < 3$ and $3 < z < 4.55$. They
ruled out a flat slope of $\sim -1$ at 5$\sigma$ and 6.5$\sigma$ levels at
these two redshift ranges, and specifically obtained a slope of $-1.6\pm0.12$
for the Ly$\alpha$ luminosity function at $z \sim 2.5$. More recently, a
wide-field ($1.43\,{\rm deg^2}$) Subaru Ly$\alpha$ survey with an unprecedented depth obtained a much
larger LAE sample of $>$3000 galaxies at $z=2.2$ \citep{Konno+2016}. This
sample yields an even steeper slope of $-1.75^{+0.10}_{-0.09}$ at $z\sim2$.
Later on, the steep slope was confirmed by another wide-field survey ($1.43\,{\rm deg^2}$) at
similar redshifts but with shallower narrowband exposures \citep{Sobral+2017}.
All those surveys indicated that there are more galaxies at faint luminosity
end, and their volume densities are much higher than those with higher
luminosities. 

 Among others, stellar mass is one of the most difficult quantities to be
measured due to their faint continuum. Usually it requires the rest-frame long
wavelength optical or near-infrared (NIR) photometry to determine a reliable
galaxy stellar mass. For an LAE at $z>2$, its rest-frame long wavelength
optical and NIR continua move to NIR or mid-IR (MIR) bands. Nonetheless, an LAE
appears to be very faint in NIR and MIR. A typical LAE at $z\sim2$ would have a
$R$-band magnitude of 25.3--25.5 magnitude \citep{Guaita+2010,Vargas+2014} and
a flat Spectral Energy Distribution (SED). Thus its NIR or MIR magnitude will
also be 25.5 magnitude or even fainter.  There are only $\sim 20\%-30\%$ of
luminous LAEs detected at 3.6\,$\mu$m, 4.5\,$\mu$m in the deep Spitzer IRAC
surveys \citep{Nilsson+2011}. It requires a large and deep coverage in NIR or
MIR to detect faint LAEs and measure their stellar masses.

Furthermore, star formation rate (SFR) and stellar mass were found to have a
tight correlation for normal star-forming galaxies, called the star formation
``main sequence'' (SFMS) \citep[e.g.,][]{Brinchmann+2004, Elbaz+2007}, which
defines a steady star formation mode.  Starburst galaxies are located above the
SFMS relation \citep[e.g.,][]{Rodighiero+2011}. At high redshifts, the SFMS
relation is derived mainly based on galaxies with stellar mass larger than
$10^{10} M_{\odot}$ \citep[e.g.,][]{Daddi+2007, Rodighiero+2011, Fang+2012,
Shivaei+2015, Shivaei+2017}, and it is often extrapolated to low mass to be
compared with LAEs. There is a debate on the locations of LAEs relative to the
SFMS relation, i.e. the existing studies found that LAEs lie above
\citep{Guaita+2013, Hagen+2014, Hagen+2016, Vargas+2014, Oteo+2015} or on
\citep{Shimakawa+2017, Kusakabe+2018} the SFMS.  It was suggested that the
inconsistent results may be caused by different survey depths or the use of
different extinction curves \citep{Shimakawa+2017, Kusakabe+2018}. 

\begin{figure}
\plottwo{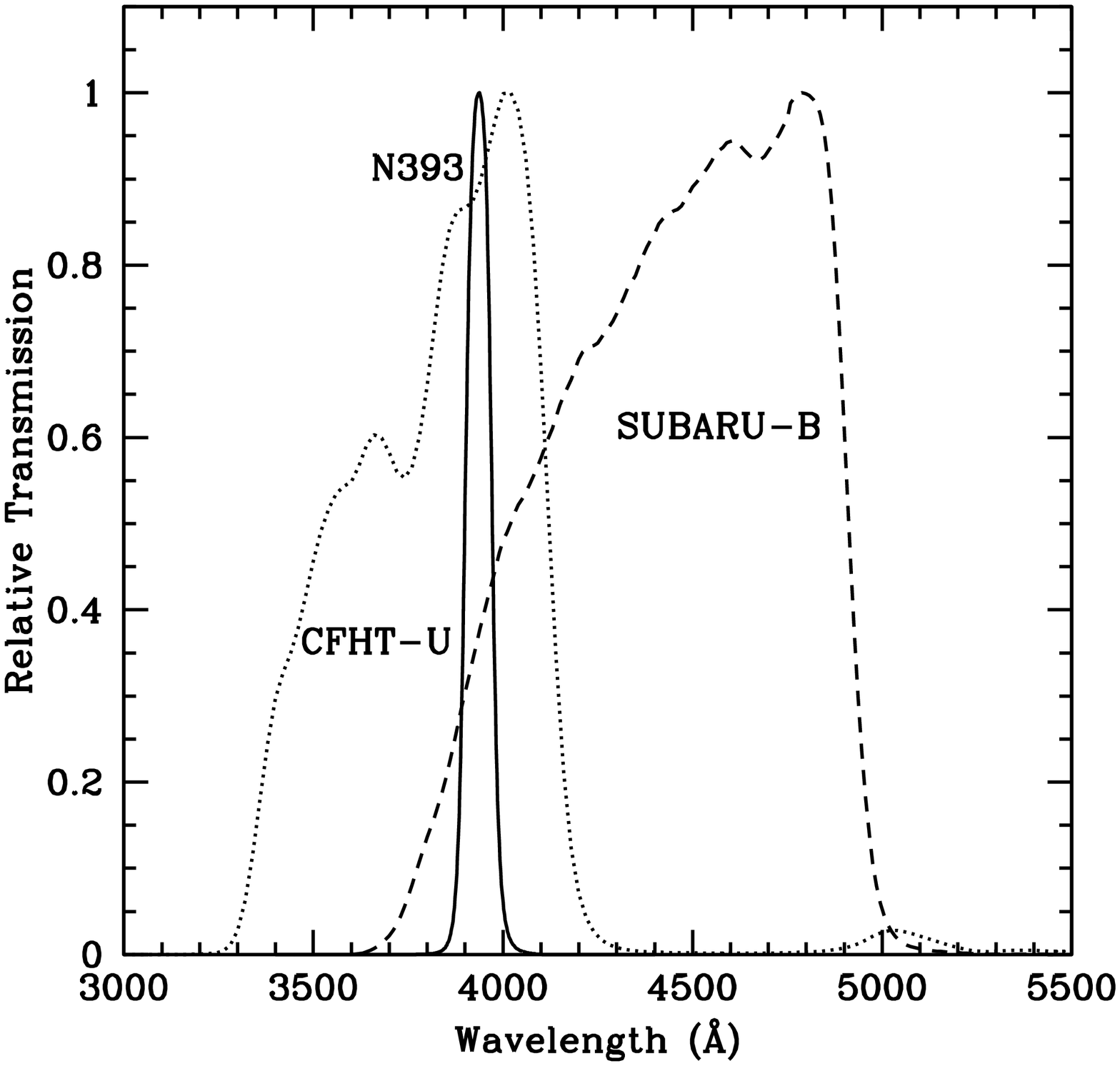}{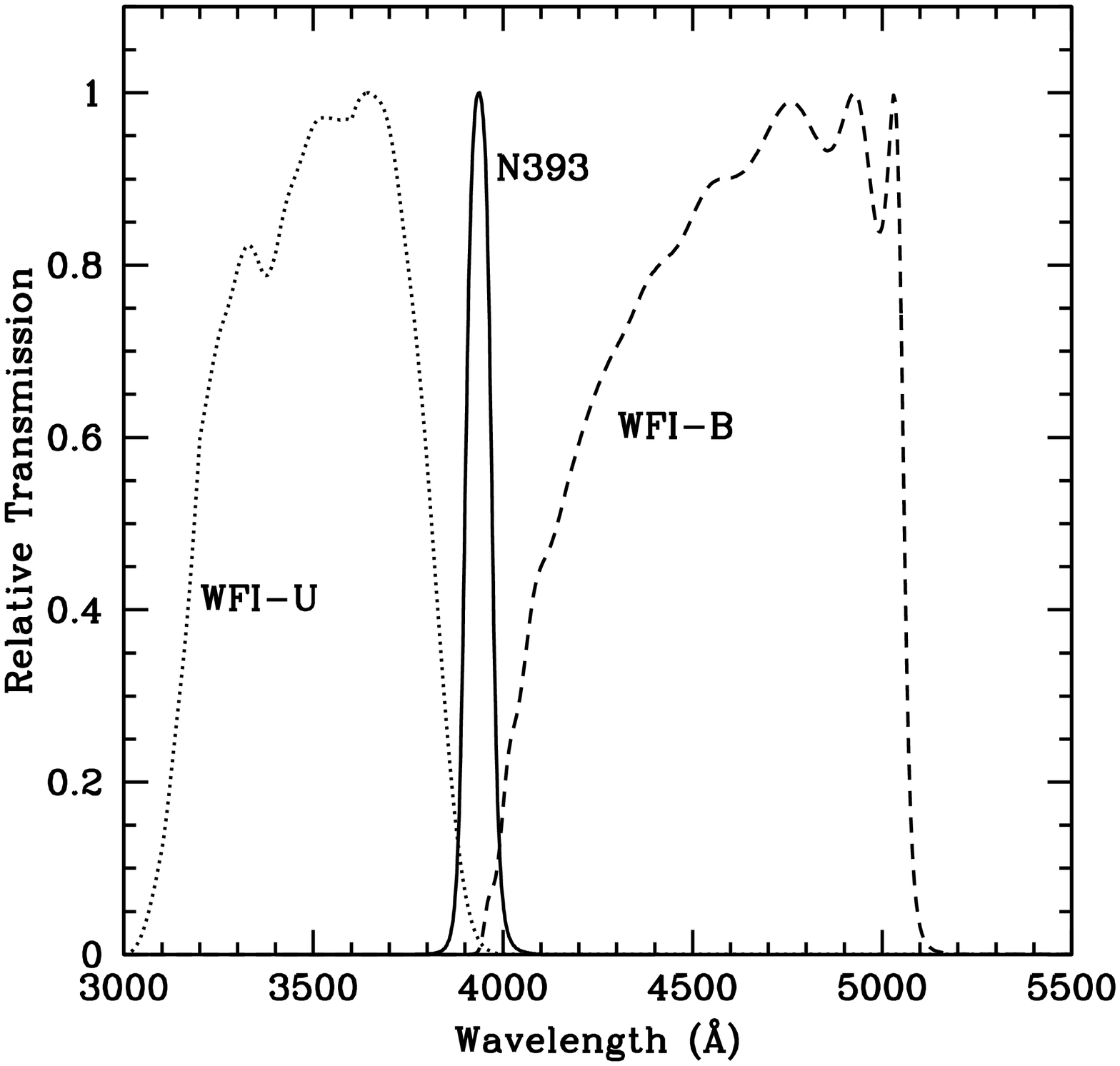}
\caption{Filter transmission curves for COSMOS (left) and ECDF-S (right). The solid line represents our customized narrowband filter N393, 
while the dotted and dashed lines represent the broad $U$ and $B$ band filters.}
\label{transmission.eps}
\end{figure}

Different survey depths of both narrowband and broadband observations would
result in different sample selection effects and sample properties. Therefore,
large and deep coverage in both narrowband and broadband is needed to provide
further independent constraints. In this paper series, we will study the
properties of LAEs using a deep narrowband-selected LAE sample with deep
multi-wavelength data. We specifically designed a customized narrowband filter
at 3928\,\AA\ with filter width of 70\,\AA\ (see Figure \ref{transmission.eps})
for detections of LAEs at $z=2.23\pm0.03$. At the same redshift, H$\alpha$
emitters can be selected using the typical NIR narrowband filter at 2.12\,$\mu$m
widely available on many telescopes \citep{Geach+2008, Sobral+2013, An+2014}.
So this filter design permits a comparison between Ly$\alpha$ and
H$\alpha$ selection for galaxy populations at $z=2.23$ \citep[e.g.,][]{Matthee+2016, An+2017, Sobral+2017}. 
We chose the COSMOS and ECDF-S
fields, where deep \textit{HST} imaging data are available including the Galaxy
Evolution from Morphologies and SEDs \citep[GEMS;][]{Rix+2004}, the Cosmic
Evolution Survey \citep[COSMOS;][]{Scoville+2007}, and the Cosmic Assembly
Near-IR Deep Extragalactic Legacy Survey \citep[CANDELS;][]{Grogin+2011,Koekemoer+2011}. 
Both fields were also covered in MIR by the \textit{Spitzer}
Extended Deep Survey \citep[SEDS;][]{Ashby+2013}. The deep \textit{Spitzer} and \textit{HST}
imaging data allow us to study the stellar masses and SFMS in this paper and UV
morphologies in a forth-coming paper (Hao et al. 2018, in preparation) .  Our
narrowband observations were carried out on Magellan II telescope with
excellent seeing condition, making it possible to study morphologies of
Ly$\alpha$ emission (Huang et al. 2018, in preparation).

This paper is structured as follows. In Section 2, we describe observations and
data reduction. Selection of LAEs is presented in Section 3.  We have our main
results in Section 4 and 5, and conclude in Section 6.  We adopt a flat
$\Lambda$CDM cosmology with $\Omega_{\rm m}=0.3$, $\Omega_{\rm \Lambda}=0.7$,
$H_{\rm 0}=70\,{\rm km \, s^{-1} Mpc^{-1}}$ and $\sigma_8=0.8$ for all
calculations, and a Salpeter Initial Mass Function \citep[IMF;][]{Salpeter1955} for stellar population analysis.

\section{OBSERVATIONS AND DATA REDUCTION}

\subsection{Observations\label{subsec:observations}}

The NB3928\,\AA\ imaging observations of the COSMOS and ECDF-S fields were
conducted with Megacam on the 6.5m Magellan II telescope.  Megacam is a
wide-field mosaic CCD camera with 9$\times$4 CCD arrays, each of which has
2048$\times$4608 pixels. The focus ratio (F/5) for Megacam on Magellan leads to
a pixel scale of 0\farcs 08/pixel, and thus an effective field-of-view of
$\sim$24\arcmin $\times$24\arcmin.  We used a binning of 2$\times$2 for a
faster readout, yielding an actual pixel scale of 0\farcs 16/pixel. The
observations were executed using dithering mode with single exposure of 15
minutes to minimize the number of saturated stars\footnote{ Stars are saturated
at a narrowband magnitude of 15.4 mag.} in each exposure. The dithering
steps vary from -49\arcsec\ to 63\arcsec\ to fill in the chip gaps. During the
narrowband observations, the seeing spans a range of 0.5\arcsec\--1.0\arcsec\
\footnote{There is only one image with a seeing larger than 1.0\arcsec\ , and
it was rejected.}, but mostly smaller than 0.7\arcsec\ . The resultant median
seeing is 0.6\arcsec\ .  After rejecting a few raw images with unusually high
sky background, we obtained total exposures of 600 and 660 minutes respectively
in the COSMOS and ECDF-S fields.  The observation parameters are summarized in
Table 1.

\subsection{Data Reduction\label{subsec:datareduction}}

The data reduction was performed with a combination of IRAF mscred package and
the customized package Megared developed in CfA/Harvard
\footnote{https://www.cfa.harvard.edu/~mashby/megacam/megacam}.  The science
images were first bias- and dark-corrected, then flat-fielded. The flat field
was generated using the twilight flat taken in this run.  We then removed the
residual pattern in each reduced frame by subtracting each frame with the mean
background frame, which was produced by stacking all science frames after
removing all objects on them.

\begin{figure}
\plottwo{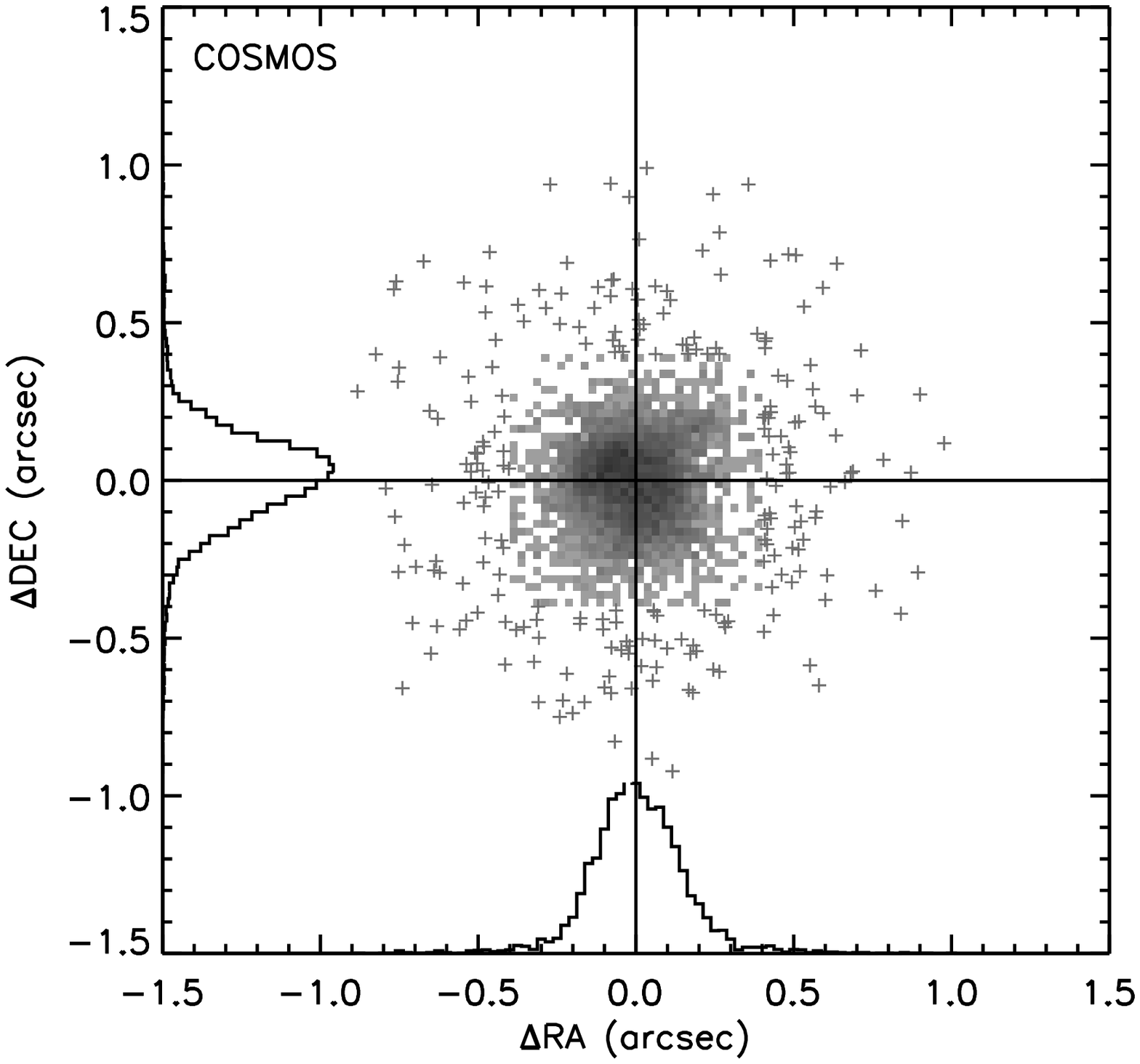}{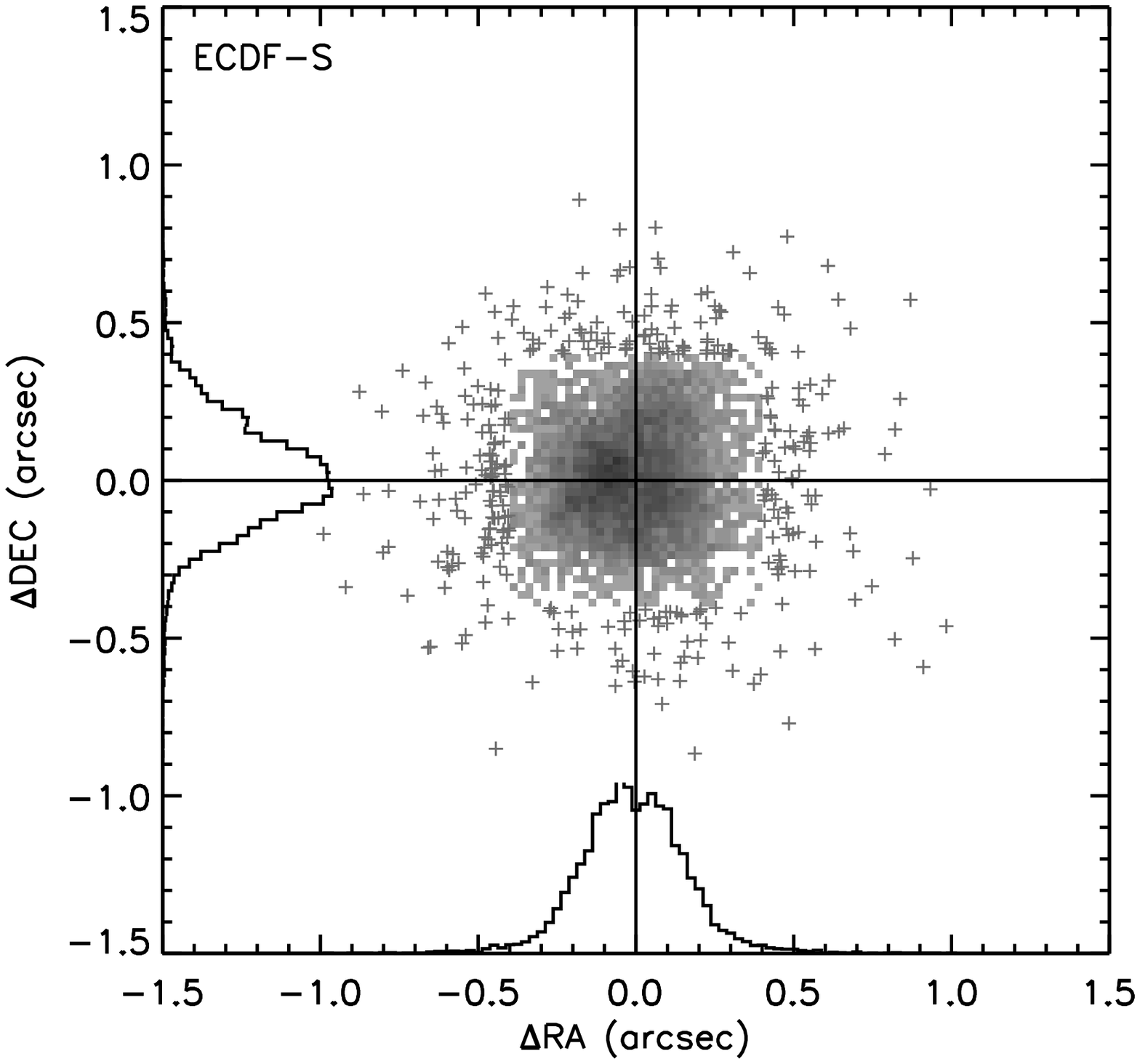}
\caption{The offset in RA and DEC between our WCS calibration and the \textit{HST}/ACS $I$-band catalog for
the COSMOS (left) and the GEMS \textit{HST}/ACS $V$-band catalog for the ECDF-S (right). }
\label{astrometryaccuracy.eps}
\end{figure}

The WCS solution for each science image needs to be refined based on the
preliminary WCS solution assigned in the observations.  We used the CfA/Harvard
developed TCL script ``megawcs'' to correct distortion and relative array
placement of each frame.  The reference positions used in this correction were
those of bright objects taken from the \textit{HST}/ACS $I$-band catalog in COSMOS \citep{Capak+2007,
Ilbert+2009} and the GEMS \textit{HST}/ACS $V$-band catalog in
ECDF-S \citep{Caldwell+2008}.  Position offsets for these objects in the refined
images and the reference catalogs are plotted in Figure
\ref{astrometryaccuracy.eps} with standard deviations of 0\farcs16 and
0\farcs17 respectively in COSMOS and ECDF-S. The accurate WCS in each image
permits optimization of the PSF in the final stacked image.

Photometric zero point in each observed frame varies due to the changes of
photometric conditions during the observations. Because we cover only one field
of view  with dithers in each field, we are able to use one set of bright stars
in each field to normalize zero point in each exposure to a reference frame. A
reference frame is one taken under the photometric condition with airmass $\sim
1$. We measured flux densities of those bright objects in each single frame and
compared them with those in the reference image. A median flux ratio was
calculated for each frame, and this ratio was used to normalize photometric
zero point of each frame to that of the reference frame before stacking.
Finally, these images were mosaiced into a single frame using Swarp software \citep{Bertin+2002}.
A coverage map was also generated accordingly in each field.  The Full Width at
Half Maximum (FWHM) for PSFs of the final stacked science images in both fields
is 0\farcs57.  The effective coverages for the two fields are nearly the same,
each $\sim$ 26\farcm9$\times$26\farcm9.

The absolute flux calibration was done using archival $U$ and $B$ band images
in COSMOS and ECDF-S. Specifically, CFHT $u^*$ and Subaru $B_{\rm J}$ from the
COSMOS archive\footnote{http://irsa.ipac.caltech.edu/data/COSMOS}
\citep{Capak+2007} and ESO MPG Wide Field Imager (WFI) $U$ and $B$ band images
from the Multi-wavelength Survey by Yale-Chile (MUSYC)
archive\footnote{http://www.astro.yale.edu/MUSYC} \citep{Taylor+2009,
Cardamone+2010} were used for the COSMOS and the ECDF-S fields respectively.
The central wavelength of our N3928\,\AA\ narrowband filter is in between the
$U$ and $B$ bands, so a linear interpolation of $U$ and $B$ band fluxes at the
central wavelength of N3928\,\AA\ can be used as the underlying continuum of
the Ly$\alpha$ emission \citep{Nilsson+2009, Guaita+2010}.  Following
\citet{Guaita+2010}, we derived the fractional contributions from $U$ and $B$
band flux densities, taking account of the central wavelengths of the
filters (see section \ref{subsec:photometry}).  Finally, by assuming that the
median color excess (see section \ref{sec:LAEselect}) of all objects is zero,
the zero point of the N3928\,\AA\ image was derived. However, as noted by
\citet{Sobral+2017}, a narrowband filter like ours covers the strong stellar
CaHK absorption feature.  A blind use of objects without considering their
spectral types could introduce problems.  So we searched for counterparts of our
narrowband-detected objects in the 3D-HST catalogs \citep{Skelton+2014} and selected star-forming
galaxies using the $UVJ$ method \citep{Williams+2009,Brammer+2011} based on the
rest-frame $U-V$ and $V-J$ colors provided by the 3D-HST catalogs. It turned out
that the narrowband zero point based on star-forming galaxies is
consistent with that obtained using all objects.  This implies that our
narrowband-detected objects are not dominated by objects with strong stellar
CaHK absorption. The 5$\sigma$ limiting magnitudes in a 3\arcsec\ diameter
aperture in the narrowband images are 25.97 and 26.02 mag for the COSMOS and
ECDF-S fields, respectively.

\subsection{Photometry\label{subsec:photometry}}

We detected objects from the narrowband image using SExtractor software
\citep{Bertin+1996}.  Objects with a minimum of 11 adjacent pixels above a
threshold of 1.5$\sigma$ per pixel were selected. The coverage map was used as
the weight image to depress spurious detections.  Aperture magnitudes with
circular apertures of diameters of 8 pixels, $\sim$ 2 FWHM of seeing disks,
were measured. We then made aperture corrections to obtain the total fluxes
using the growth curve derived from bright stars. We note that such aperture
corrections are made by assuming that the objects are point-like sources.  Such
an assumption is reasonable for the vast majority of our LAEs (Huang et al.
2018, in preparation). We note that MAG\_AUTO from SExtractor is often used to
measure the total Ly$\alpha$ fluxes in the literature. However, the fluxes
measured by MAG\_AUTO are dependent on the survey depth and are biased
measurements for faint objects near the detection limits, as noted by
\citet{Matthee+2016} and \citet{Konno+2014}. This was also seen in our
data. We found that the aperture-corrected fluxes under-estimate the
total fluxes probed by MAG\_AUTO only for bright LAEs, while the MAG\_AUTO
fluxes for LAEs with $NB > 25 $mag are consistent with the aperture-corrected
fluxes on average but with larger photometric errors. Therefore, we decided to
use the aperture-corrected flux as a measure of the total flux in the following
analysis. The exception is that MAG\_AUTO was also used in the construction of
Ly$\alpha$ luminosity functions (See section \ref{sec:LF}) for the purpose of
comparisons with studies in the literature.

Objects with signal-to-noise ratio less than 5 in the narrowband photometry were
excluded. But objects fainter than the 5$\sigma$ limiting magnitude were not
removed from the catalog.  We masked out saturated stars and high noise area
(with exposure time $\leq$40 minutes) in the narrowband images. The objects
located in the masked area were accordingly removed from the final catalog. The
final narrowband photometry catalogs consist of 25,756 and 27,946 objects,
covering effective area of 602.05 and 612.75 arcmin$^2$ in the COSMOS and
ECDF-S fields, respectively.

For the calculation of the narrow-to-broad band color excess, we re-binned the
$U$ and $B$ band images to match the pixel scale of the narrowband image in each
field and then performed aperture photometry on the broadband images using
SExtractor software in ``dual image" mode. The narrowband image was used as the
detection image and the broadband images were taken as measurement images. We
used the same set-up in the ``dual image" mode as that used for the source
detection, as listed above.  The seeing FWHM of $U$ and $B$ band images are
0\farcs 87 and 0\farcs 94 for COSMOS and 1\farcs 03 and 1\farcs 02 for ECDF-S.
Circular aperture photometry with diameters of 12 (14) pixels was carried out
on both $U$ and $B$ band images for the COSMOS (ECDF-S) field.  For some
objects detected in the narrowband, the measured broadband fluxes have
signal-to-noise ratio less than 2. We then used $2\sigma$ flux limits 
for these objects. Similar to the narrowband case, aperture
corrections were subsequently made to measure the total fluxes using the growth
curves obtained from bright stars in $U$ and $B$ bands. Aperture corrected
magnitudes were used to calculate the narrow-to-broad band color excess.

 After the photometry in both narrowband and its adjacent broadbands was
obtained, the underlying continuum of the Ly$\alpha$ line, denoted by
$f_{\lambda,UB,con}$, the Ly$\alpha$ equivalent width (EW) and the Ly$\alpha$ line flux
could be calculated.  Since the broadband observations of the two fields used
different filters, we adopted different equations to calculate these
quantities. As mentioned in Section \ref{subsec:datareduction}, we used a
linear combination of $U$ and $B$ band flux densities to measure the continuum
with the central wavelengths of the filters taken into account. 
Specifically, the interpolated $U$ and $B$ band flux densities at the
narrowband central wavelength is derived via the linear interpolation formula
\begin{equation}
\frac{f_{\lambda,UB}-f_{\lambda,U}}{\lambda_{NB}-\lambda_U} = \frac{f_{\lambda,B}-f_{\lambda,U}}{\lambda_B-\lambda_U},
\end{equation}
where $f_{\lambda,UB}$ is the interpolated $U$ and $B$ band flux densities at the narrowband central wavelength,
$f_{\lambda,U}$ and $f_{\lambda,B}$ are the $U$ and $B$ band flux densities, while $\lambda_{NB}$, $\lambda_U$ and $\lambda_B$
are the central wavelengths of the narrowband, $U$ and $B$ band filters, respectively.
For COSMOS, $f_{\lambda,UB}=0.80f_{\lambda,U}+0.20f_{\lambda,B}$,
while for ECDF-S, $f_{\lambda,UB}=0.57f_{\lambda,U}+0.43f_{\lambda,B}$.
For the COSMOS field, we note that the CFHT $U$ band filter includes the
Ly$\alpha$ line (See the left panel of Figure \ref{transmission.eps}).
Therefore, we need to remove the Ly$\alpha$ emission from the observed $U$ band
flux density before it is used to calculate the underlying continuum. For LAEs
in COSMOS, the equation used to calculate the continuum of the Ly$\alpha$ line
is:
\begin{equation}
f_{\lambda,UB,con}=\frac {f_{\lambda,UB}-0.80 f_{\lambda,N393}\frac{\Delta \lambda_{NB}}{\Delta \lambda_{U}}}{1-0.80\frac{\Delta \lambda_{NB}}{\Delta \lambda_{U}}},
\end{equation}
where $f_{\lambda,N393}$ is the narrowband flux density, while $\Delta \lambda_{NB}$ and $\Delta \lambda_{U}$ are the bandwidth of the narrowband and $U$ band filters, respectively.
Accordingly, the EW for LAEs in the COSMOS field can be derived using the following equation:
\begin{equation}
EW_{\rm obs}=\frac {f_{\lambda ,N393}-f_{\lambda ,UB}}{f_{\lambda ,UB}-f_{\lambda ,N393} \frac {0.80 \Delta \lambda_{NB}}{\Delta \lambda_{U}}}\, \Delta \lambda_{NB},
\end{equation}
where $EW_{\rm obs}$ is the observed EW that is related to the rest-frame EW by
$EW_{\rm obs}=(1+z)EW_{\rm rest}$.
The Ly$\alpha$ flux is obtained as follows:
\begin{equation}
F({\rm Ly\alpha})=\frac {f_{\lambda ,N393}-f_{\lambda ,UB}}{1-0.80\frac{\Delta \lambda_{NB}}{\Delta \lambda_{U}}}\, \Delta \lambda_{NB} .
\label{eq:cosmoslyaflux}
\end{equation}
For the case of the ECDF-S field, the calculations of $f_{\lambda,UB,con}$, Ly$\alpha$ EW and Ly$\alpha$ line
flux are simpler since the Ly$\alpha$ line is not included in the broadband filters.
So for LAEs in ECDF-S,
\begin{equation}
f_{\lambda,UB,con}=f_{\lambda,UB},
\end{equation}
\begin{equation}
EW_{\rm obs}=\frac {f_{\lambda ,N393}-f_{\lambda ,UB}}{f_{\lambda ,UB}}\, \Delta \lambda_{NB},
\end{equation}
and the Ly$\alpha$ flux is derived using the following equation:
\begin{equation}
F({\rm Ly\alpha})= (f_{\lambda ,N393} - f_{\lambda ,UB})\, \Delta \lambda_{NB} .
\label{eq:ecdfslyaflux}
\end{equation}

\subsection{Survey Completeness\label{subsec:completeness}}

It is essential to determine the completeness of the narrowband surveys.  The
Monte Carlo simulations were employed to measure the narrowband survey
completeness in both fields respectively. We first selected several tens of
bright stars (SExtractor parameter CLASS\_STAR $\ge 0.95$) with the
aperture-corrected narrowband magnitudes $19 \leq NB < 22$ mag in both fields,
scaled down their flux densities, and put back in random locations in their
original images.  SExtractor with the same parameter set was run again on the
images with artificial objects. This simulation was performed several hundred
times to achieve adequate statistics in each magnitude bin.  Completeness of
the narrowband detections in both fields was measured as the artificial object
recovery rate in each magnitude bin, shown as the solid data points in
Figure \ref{comparecompleteness_usestackimagemag_auto_usestarapercor.eps}.  The
90\% and 50\% completeness limits are 25.50 mag and 25.99 mag for the COSMOS
field and 25.78 mag and 26.20 mag for the ECDF-S field.

\section{SELECTION OF LAE CANDIDATES\label{sec:LAEselect}} 
 
Selection of LAEs is generally based on their emission line EW. Practically the
narrowband-to-broadband color excess (i.e., magnitude difference) can be used
as a proxy. We use UB as magnitude for the interpolated $U$ and $B$ bands flux
density, and NB as narrowband magnitude.  The median NB-UB should be zero for
galaxies with no line shifting to the band.  When a line shifts to the
narrowband, the NB-UB appears ``blue''. We plot NB-UB against NB in Figure
\ref{colorexcessplot.eps}. The scatters in NB-UB come from the NB, $U$, and $B$
band photometry uncertainties. We measured the rms scatters $\sigma_{NB-UB}$ in
the NB-UB distributions as a function of NB magnitude and used
$3\sigma_{NB-UB}$ as the threshold to select narrowband excess sources as LAE
candidates. Due to the different depths in broadband photometry in these two fields,
EWs calculated using $3\sigma_{NB-UB}$ are slightly different,  $\sim20$\,\AA\
in COSMOS and $\sim30$\,\AA\ in ECDF-S respectively.  If a selection
criterion of ${\rm EW} \geq  30$\,\AA\ is used for COSMOS, the number densities
and main results do not change significantly (see below for details). This
selection left us with 212 LAEs in COSMOS and 263 LAEs in ECDF-S shown in
Figure \ref{colorexcessplot.eps}.

The narrowband-excess sample yielded by the single NB-UB selection also
includes objects with emission lines other than Ly$\alpha$ line
red-shifting to our narrowband filter waveband.  For example,
[O\,{\scriptsize II}] at z$\sim 0.05$; Al\,{\scriptsize III} at z$\sim 1.1$;
C{\scriptsize III}] at $z\sim 1.06$; C\,{\scriptsize IV} at z$\sim 1.5$; and
Si\,{\scriptsize IV} + O\,{\scriptsize IV} at z$\sim 1.8$ etc. Among these
lines, only [O\,{\scriptsize II}] appears in a normal galaxy spectrum, while
the remaining high ionization lines appear in active galactic nuclei (AGNs)
spectra. The [O\,{\scriptsize II}] emitters are at such a low redshift that
they occupy too small volume to actually contribute to the selected sample.
Those high ionization line emitters were identified using X-ray catalogs
\citep{Civano+2012, Luo+2008, Lehmer+2005, Virani+2006} and spectroscopic
information from NED\footnote{http://nedwww.ipac.caltech.edu/}.  We identified
16 broad-line AGNs and one candidate AGN with X-ray emission at $z=1.694$
\citep{Salvato+2011} in COSMOS, and 7 broad-line AGNs in ECDF-S in the
narrowband-excess sample. Four of the broad-line AGNs in COSMOS
\citep{Trump+2009,Civano+2012} and 2 in ECDF-S
\citep{Treister+2009,Silverman+2010} have spectroscopic redshifts of
$\sim2.23$.  For the study of the Ly$\alpha$ luminosity functions in Section
\ref{sec:LF}, we include these $z\sim2.23$ broad-line AGNs and reject all the other AGNs.
But for the study of star formation properties of LAEs, all broad-line AGNs are removed.
The other possible contamination sources are Lyman-Break Galaxies (LBGs) at
$z\sim3$. At $z\sim3$, both the narrowband and broad $B$ band sample the
continuum at rest-frame wavelength longer than 912\AA\ while the $U$-band
samples the rest-frame flux shorter than 912\AA. The break makes interpolation
between $U$ and $B$ band artificially low and NB-UB appears to be excessive.  We
simply use NB-$B$ color to reject $z\sim3$ LBGs. The reliability of this criterion was
tested by LBGs selected by the classical LBG technique.  Specifically, we apply
the criteria proposed by \citet{Alvarez-Marquez+2016} to select $z\sim3$ LBGs in our
COSMOS field using the COSMOS public archival catalogue and the criteria
adopted by \citet{Hildebrandt+2005} to select $z\sim3$ LBGs in our ECDF-S field using
the MUSYC public catalogue.  As shown in Figure \ref{LBGs.eps}, LBGs mostly have
${NB-B} \ge 0$ and LAEs have ${NB-B}<0$, although a minority of LBGs have
${NB-B}<0$. A more strict criterion of ${NB-B} \ge -0.3$ would not change our
results significantly. This criterion only identifies 5 $z\sim3$ LBGs in COSMOS that
were rejected from the sample. The final sample consists of 194 LAE candidates
(including 4 AGNs) in COSMOS and 258 LAE candidates (including 2 AGNs) in
ECDF-S, respectively.

Both EW and narrowband flux limit have selection effects in a
narrowband-excess-selected LAE sample. Figure \ref{LAE_Lyalumi_EWrest.eps}
shows that our sample does not include LAEs with low Ly$\alpha$ luminosity and
high EW.  This results from the narrowband detection limits. An LAE with low
Ly$\alpha$ luminosity and high EW implies a low continuum, therefore its
narrowband flux, i.e., the sum of the emission line flux and the continuum, is
too low to be detected in the narrowband imaging \citep{Guaita+2010}.  From
Figure \ref{LAE_Lyalumi_EWrest.eps}, we also see that a larger number of faint
LAEs were selected in ECDF-S than in COSMOS because of the deeper narrowband
exposure in the ECDF-S field. The different selection criteria in EW
employed by these two fields are also clearly seen in this figure. There are 31
LAEs with rest-frame EW between 20\AA\ and  30\AA\ in COSMOS, $\sim16\%$ of the
LAE sample in COSMOS. However, the inclusion of these LAEs compared to a
selection criterion of ${\rm EW} \geq 30$\AA\ does not change the number
densities per luminosity bin significantly, due to the wide spread of these
LAEs in the Ly$\alpha$ luminosity as shown in Figure
\ref{LAE_Lyalumi_EWrest.eps}. The changes are mostly within the 1$\sigma$
Poisson noises.

\section{LAE Number Counts and Ly$\alpha$ Luminosity Function \label{sec:LF}}

\begin{figure}
\plotone{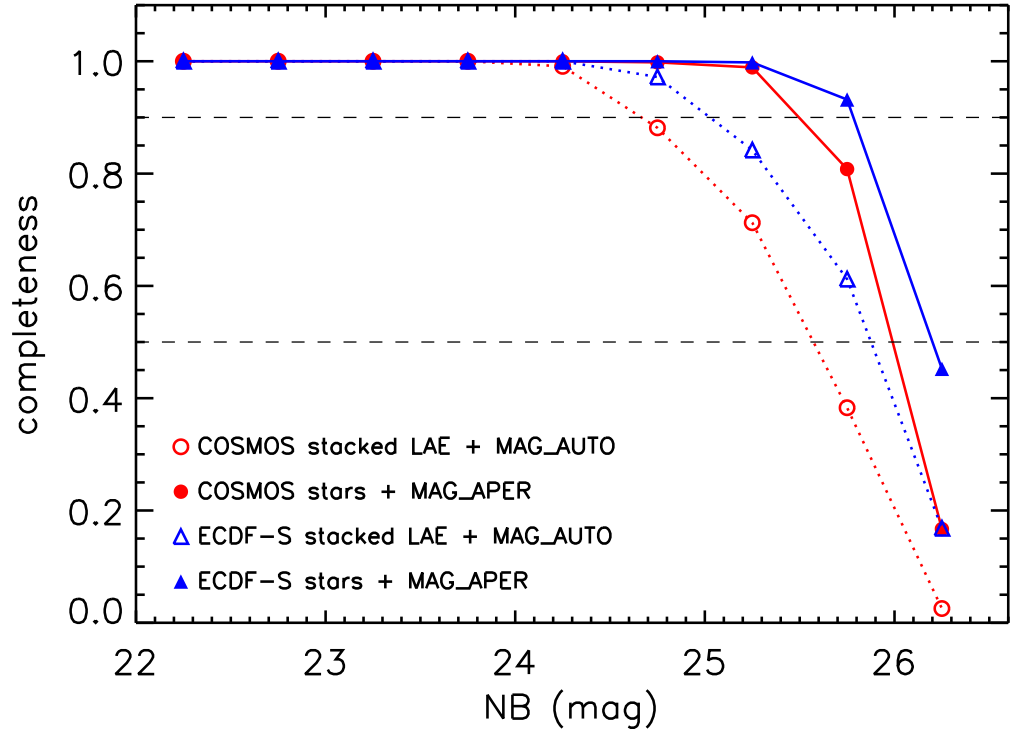}
\caption{Completeness curves for COSMOS (red points) and ECDF-S (blue points).
The solid points represent the completeness curves obtained using bright stars and aperture-corrected magnitudes,
while the the open symbols indicate the completeness curves based on the stacked narrowband image of our LAEs and MAG\_AUTO magnitudes.
}
\label{comparecompleteness_usestackimagemag_auto_usestarapercor.eps}
\end{figure}

\begin{figure}
\plotone{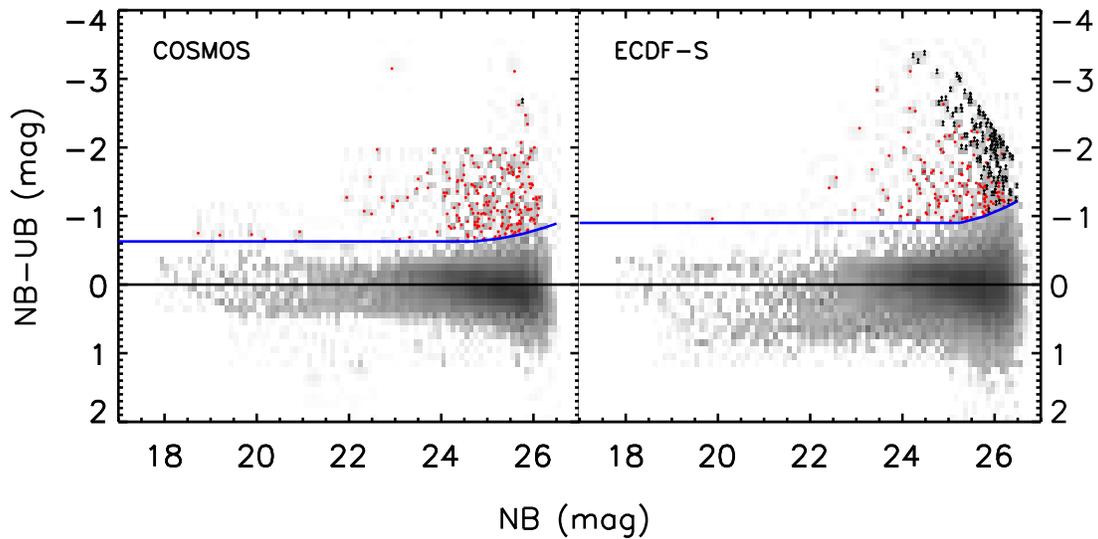}
\caption{Narrowband color excess as a function of narrowband magnitude for the COSMOS (left) and ECDF-S (right) fields.
The gray-scale represents the number density of the N393-detected objects. 
The black solid line indicates zero line emission or absorption, while the blue solid curve represents the 3$\sigma$ rms scatter selection criteria of LAE candidates
as a function of narrowband magnitude. The red dots
show the 3$\sigma$ selected LAE candidates and the upper arrows denote the LAE candidates that were not detected at either $U$ or $B$ band
at a 2$\sigma$ level.}
\label{colorexcessplot.eps}
\end{figure}

\begin{figure}
\plotone{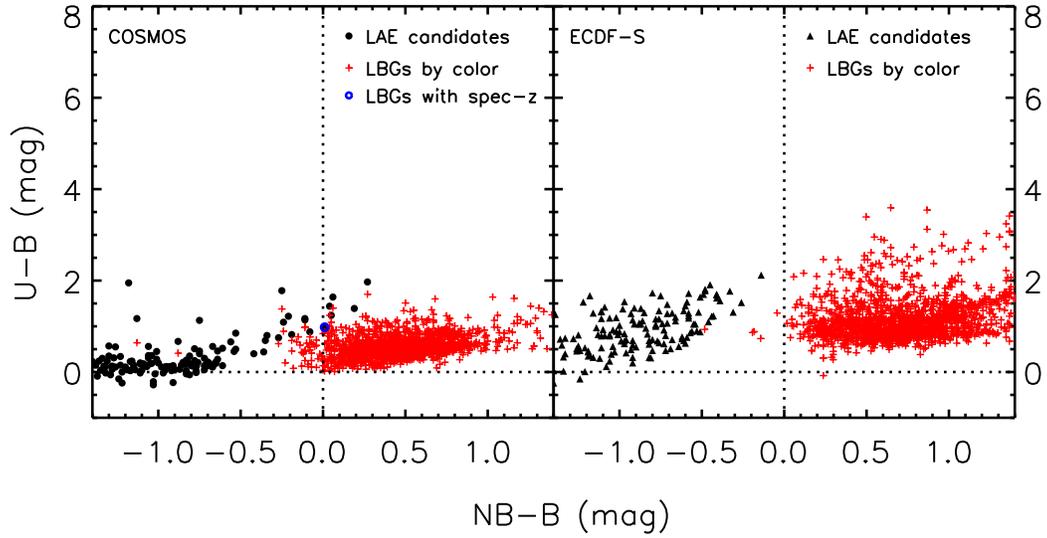}
\caption{Selection of $z\sim3$ LBGs by $U$-$B$ vs. NB-$B$ colors for the COSMOS (left) and ECDF-S (right) fields. 
The selection criterion $NB-B > 0$ is plotted as a dotted line. The filled circles in the left panel
and the filled triangles in the right panel represent LAE candidates. The red crosses are $z\sim3$ LBGs selected
by commonly used broadband LBG technique \citep{Alvarez-Marquez+2016,Hildebrandt+2005} and the blue open circle denotes a $z\sim3$ LBG that has already
been confirmed by spectroscopic observation \citep{Lilly+2007}.}
\label{LBGs.eps}
\end{figure}

\begin{figure}
\plotone{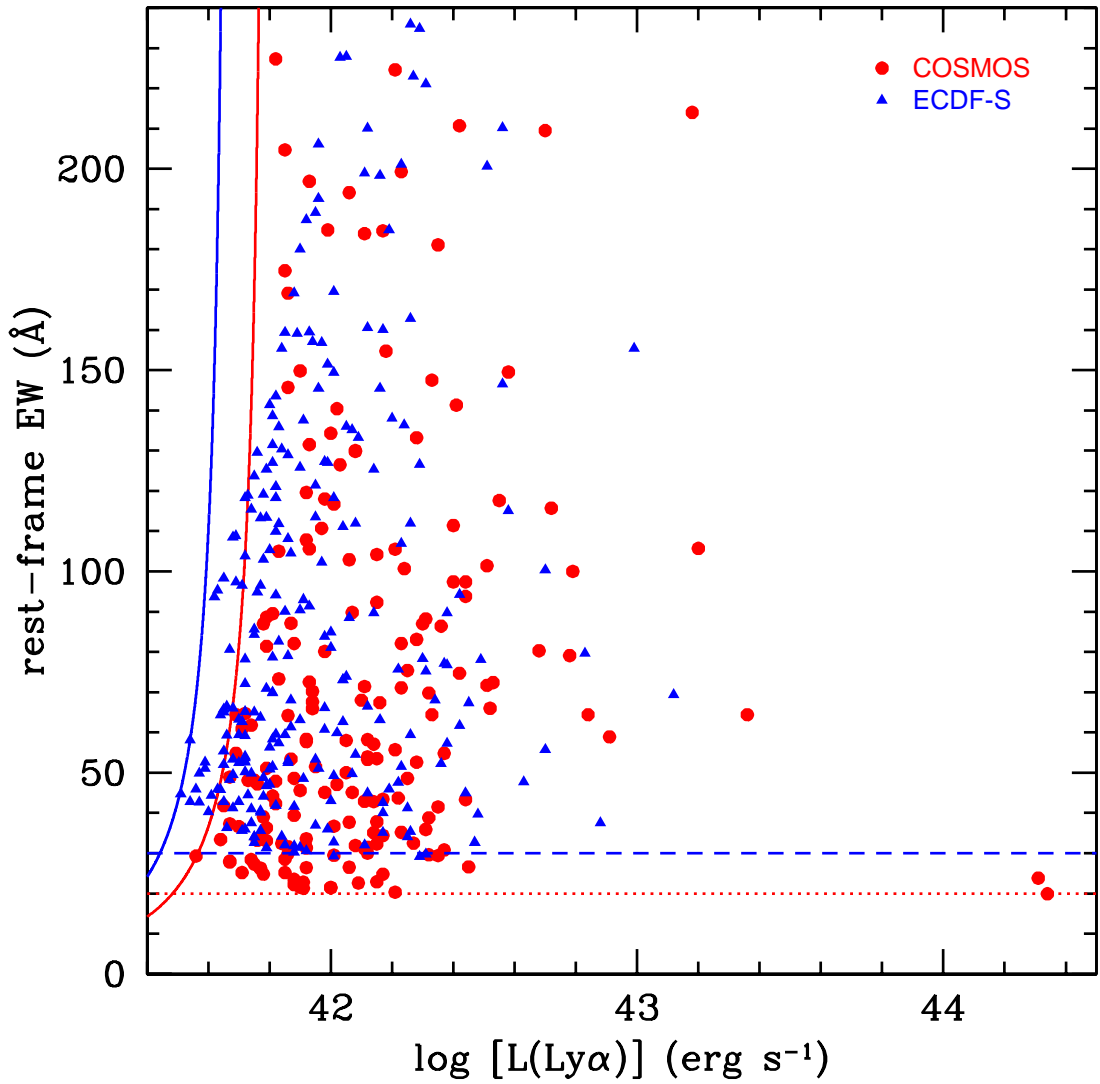}
\caption{Selection effects in the COSMOS (red circles) and ECDF-S (blue triangles) fields shown in the rest-frame EW 
versus logarithmic Ly$\alpha$ luminosity plot.
The red and blue horizontal lines indicate the color excess selection threshold for the two fields that are equivalent to Ly$\alpha$ EW of 20\,\AA\ and 30\,\AA\ for 
the COSMOS and ECDF-S fields, respectively. The red and blue solid curves correspond to the faintest magnitude in the COSMOS and ECDF-S LAEs samples. 
}
\label{LAE_Lyalumi_EWrest.eps}
\end{figure}

\begin{figure}
\plotone{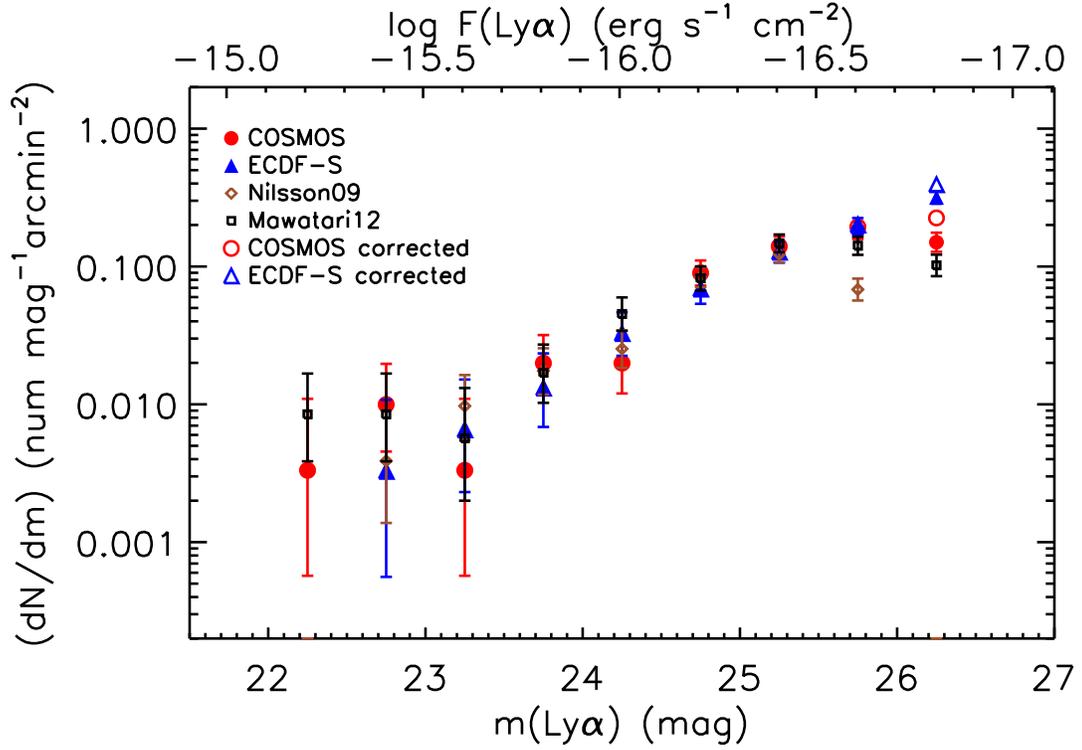}
\caption{Number counts of LAEs in terms of Ly$\alpha$ magnitude/flux as derived using 
equation (\ref{eq:cosmoslyaflux})  in COSMOS (red filled circles) 
and equation (\ref{eq:ecdfslyaflux}) in ECDF-S (blue filled triangles). The corresponding open symbols represent 
the completeness-corrected values. LAEs at $z\sim2.25$ in \citet{Nilsson+2009} and
LAEs at $z\sim2.4$ in \citet{Mawatari+2012} are plotted in brown diamonds and black squares
for comparison.}
\label{LAEnumbercounts.eps}
\end{figure}

Measurement of galaxy number counts is a direct way of estimating depth of an
imaging survey. We use Ly$\alpha$ magnitude/flux (with continuum
subtracted, see equations (\ref{eq:cosmoslyaflux}) and (\ref{eq:ecdfslyaflux})
) to perform the analysis. The Ly$\alpha$ magnitude $m({\rm Ly\alpha})$
is linked to the Ly$\alpha$ flux via the equation $m({\rm Ly\alpha})=-2.5
{\rm log} (\frac{F({\rm Ly\alpha})}{\Delta \lambda_{NB}} \frac{\lambda_{NB}^2}{c})+m_0$, where
c is the speed of light and $m_0$ is the zero-point of the AB magnitude
system. The LAE number counts of the COSMOS and ECDF-S fields are listed in
Table 2. In Figure \ref{LAEnumbercounts.eps}, we plot our counts against those
in previous narrowband surveys with publicly available catalogs at similar
redshifts. It shows that our counts reach a limiting flux density of
1.58$\times$10$^{-17}$ erg\,s$^{-1}$\,cm$^{-2}$. It is clear from Figure
\ref{LAEnumbercounts.eps} that most surveys have consistent number counts up to
$F{(\rm Ly\alpha)}$=2.5$\times$10$^{-17}$ erg\,s$^{-1}$\,cm$^{-2}$, but at the
faintest luminosity bin of $F{(\rm Ly\alpha)}$=1.58$\times$10$^{-17}$
erg\,s$^{-1}$\,cm$^{-2}$, our survey has the highest counts comparing with
previous surveys \citep{Nilsson+2009, Mawatari+2012}.  Note that the narrowband
imaging completeness corrected LAE number counts of the COSMOS and ECDF-S
fields are different at the faintest end. This may be caused by different
narrowband depths in the two fields and potential uncertainties associated with the
large completeness correction in COSMOS at the faintest magnitude bin, as shown
in Figure \ref{LAE_Lyalumi_EWrest.eps} and Figure
\ref{comparecompleteness_usestackimagemag_auto_usestarapercor.eps}
respectively.

With the deep LAE sample, we estimated Ly$\alpha$ luminosity function with the
$<1/{\rm V_{max}}>$ method \citep{Shimasaku+2006, Ouchi+2008, Gronwall+2007,
Hu+2010, Konno+2016}.  For a  boxcar shape narrowband filter, it would be
straightforward to derive the luminosity function by simply dividing the
observed incompleteness-corrected number counts by their effective volume,
i.e., the classical method \citep[e.g.,][]{Shimasaku+2006, Ouchi+2008}.
However, in reality narrowband filter transmission curve does not have a boxcar
shape, and the classical method suffers from some uncertainties, as noted by
\citet{Ouchi+2008}.  \citet{Ouchi+2008} showed that these uncertainties are not
important and/or cancel each other, but this may not apply to our samples due
to the different filter shapes and survey properties. Unfortunately, the sample
sizes and the narrow coverage in Ly$\alpha$ luminosity do not allow us to
perform a sophisticated simulation like \citet{Ouchi+2008}. So we assume the
potential uncertainties caused by our non-boxcar filter profile cancel each
other too.

\begin{figure}
\plotone{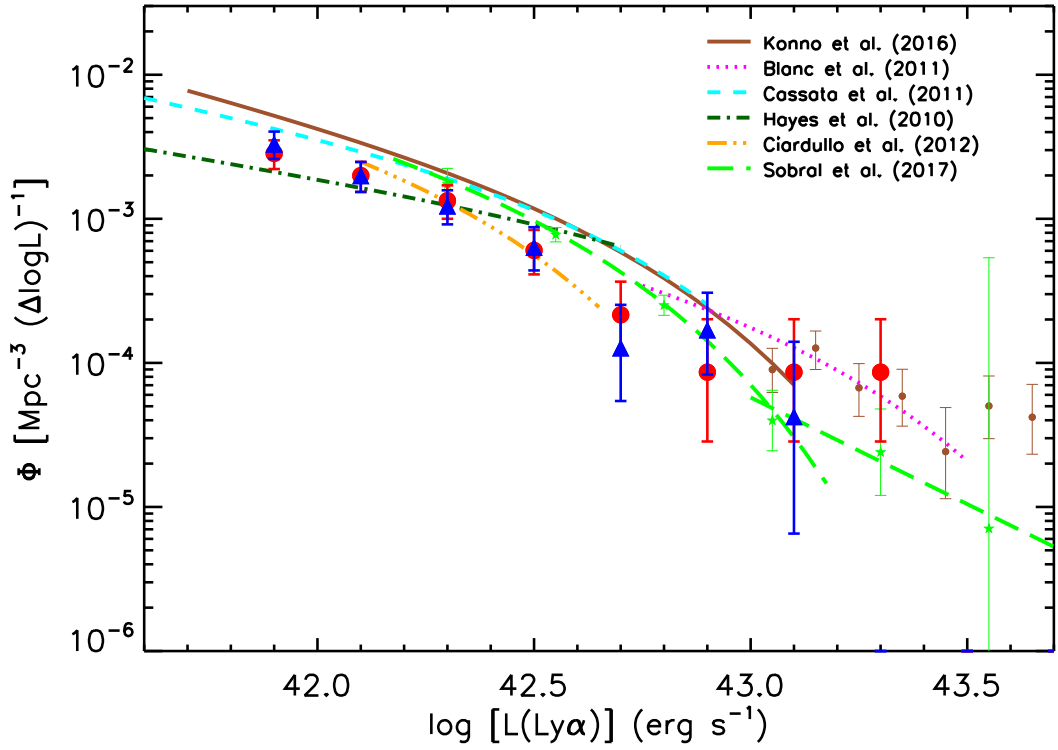}
\caption{Comparisons of $z\sim2$ Ly$\alpha$ luminosity functions in this work and those in the literature.
The big data points represent our observed luminosity functions derived using aperture-corrected fluxes and stars-based 
completeness curves for the COSMOS (red circles) and ECDF-S (blue triangles) fields. The color-coded curves show 
luminosity functions by different groups, 
as labeled at the upper-right corner. Data points from \citet{Sobral+2017} and the data points with $L(Ly\alpha) > 10^{43}\,
{\rm erg\,s^{-1}}$ from \citet{Konno+2016} are plotted
as small green stars and small brown filled circles, respectively.
}
\label{compareLFshapeliterature.eps}
\end{figure}

Under the assumption that the redshift range corresponds to the FWHM of the
narrowband filter, the effective co-moving volumes probed by our narrowband
surveys for COSMOS and ECDF-S fields are $1.16\times 10^5 \, {\rm Mpc}^3$ and
$1.18\times 10^5 \, {\rm Mpc}^3$, respectively.  The derived Ly$\alpha$
luminosity functions are provided in Table 3 and shown in Figure
\ref{compareLFshapeliterature.eps}. The completeness at $L({\rm Ly\alpha}) <
10^{41.8}\, {erg\,s^{-1}}$ is lower than 20\%. We do not use LAEs below
$10^{41.8}\, {erg\,s^{-1}}$ in the estimation of the luminosity functions. The
errors include uncertainties from both Poisson noise \citep{Gehrels1986} and
cosmic variance.  Following the method in \citet{Konno+2016}, the cosmic
variance uncertainty is obtained from the bias factor of our LAEs derived in
Section \ref{subsec:clustering} and the dark matter density fluctuation in a
sphere with a radius of $\sim 30\,{\rm Mpc}$ (corresponding to our survey
volume) at redshift 2.23.  For both fields, the resultant cosmic variance
uncertainty is 17.7\%.

As can be seen in Figure \ref{compareLFshapeliterature.eps}, our Ly$\alpha$
luminosity functions of the two fields agree with each other and are generally
consistent with those obtained in previous surveys \citep{Hayes+2010,
Blanc+2011, Cassata+2011, Ciardullo+2012, Konno+2016}. We specifically compare
our results with the recent work by \citet{Konno+2016} and \citet{Sobral+2017}.
For a better comparison, apart from the best-fit luminosity functions, data
points from \citet{Sobral+2017} and data points with $L({\rm Ly\alpha}) > 10^{43}\,
{\rm erg\,s^{-1}}$ from \citet{Konno+2016} are also plotted in Figure
\ref{compareLFshapeliterature.eps}. Both surveys observed an effective area of
$1.43\, {\rm deg}^2$ and probe a co-moving volume of $1.32 \times 10^6\, {\rm
Mpc}^3$ \citep{Konno+2016} and $7.3 \times 10^5\, {\rm Mpc}^3$
\citep{Sobral+2017}, respectively.  The limiting Ly$\alpha$ luminosities are
$10^{41.7}\, {\rm erg\,s^{-1}}$ for \citet{Konno+2016} and $10^{42.3}\, {\rm
erg\,s^{-1}}$ for \citet{Sobral+2017}, respectively. Although these surveys
probe much larger areas and co-moving volumes than ours, we reach a comparable depth of
Ly$\alpha$ luminosity limit of $10^{41.8}\, {\rm erg\,s^{-1}}$ with
\citet{Konno+2016} in the luminosity functions.  \citet{Sobral+2017} has made
extensive comparisons with \citet{Konno+2016}.  They showed that when the same
EW threshold was used and no contaminants were removed, their Ly$\alpha$
luminosity function would be perfectly consistent with \citet{Konno+2016}. Even
without matching the selection criteria, their results are claimed to be in good
agreement with \citet{Konno+2016} over most luminosities.  Our luminosity
functions roughly follow the trend of them, but the volume densities are
systematically lower at $L({\rm Ly\alpha}) < 10^{43}\, {\rm erg\,s^{-1}}$. The possible
causes for the differences are two-folds. One is the measurement of the total
luminosity.  We used aperture-corrected fluxes to probe the total flux while
\citet{Konno+2016} used MAG\_AUTO and \citet{Sobral+2017} used a larger
aperture to measure the total flux.  The other is the completeness correction.
We used bright stars to evaluate the detection completeness and assumed effects
caused by our non-boxcar filter profile cancel each other, similar to
\citet{Konno+2016}. But \citet{Sobral+2017} performed more corrections,
including both selection completeness and filter profile biases.

\begin{figure}
\plotone{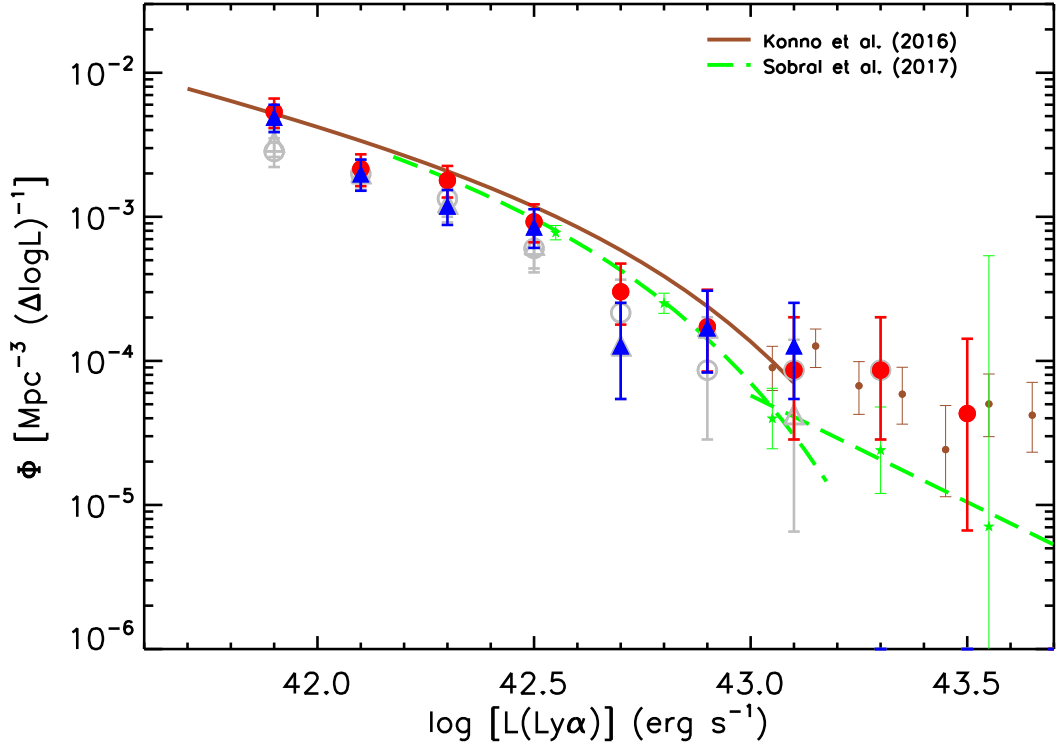}
\caption{Comparisons of the Ly$\alpha$ luminosity functions based on MAG\_AUTO and completeness
derived using the reconstructed LAE image (red solid circles for COSMOS and blue solid 
triangles for ECDF-S) with the Ly$\alpha$ luminosity functions based on the aperture-corrected fluxes and 
stars-based completeness curves (gray open circles for COSMOS and gray open triangles for ECDF-S) and those in
\citet{Konno+2016} and \citet{Sobral+2017}.
The best-fit Ly$\alpha$ luminosity functions from \citet{Konno+2016} and \citet{Sobral+2017} are 
shown in color-coded curves, as denoted at the upper-right corner.
Data points from \citet{Sobral+2017} and the data points with $L({\rm Ly\alpha}) > 10^{43}\,
{\rm erg\,s^{-1}}$ from \citet{Konno+2016} are represented
by green stars and small brown points, respectively.
}
\label{f10.eps}
\end{figure}

To understand our systematically lower number densities than
\citet{Konno+2016} and \citet{Sobral+2017}, we derive Ly$\alpha$ luminosity
functions using MAG\_AUTO as a measure of the total flux and completeness
curves built upon a reconstructed LAE narrowband image.  The reconstructed LAE
narrowband image is obtained by stacking the narrowband images of our LAEs. The
approach used to construct the completeness curves is similar to that using
bright stars in Section \ref{subsec:completeness}. The difference is that the
reconstructed and flux-scaled LAE narrowband images are used as the input fake
narrowband images and MAG\_AUTO is used to measure the flux of the fake
objects. The stacked LAE for each field has a FWHM of $\sim0.7$\arcsec\ ,
slightly broader than the PSF. It is the requirement of the recovery (within
3$\sigma$) of the input MAG\_AUTO that makes the detection completeness
different from the ones using bright stars and aperture-corrected magnitudes,
as shown in Figure
\ref{comparecompleteness_usestackimagemag_auto_usestarapercor.eps}. The
resultant Ly$\alpha$ luminosity functions are added to Table 3 and shown in
Figure \ref{f10.eps}. The luminosity functions based on aperture-corrected
fluxes and completeness curves derived with stars are overplotted as gray
open points for comparison. As can be seen, our results based on MAG\_AUTO
and completeness from the reconstructed LAE image agree well with the
Ly$\alpha$ luminosity function of \citet{Sobral+2017}, although the number
densities are still slightly lower than \citet{Konno+2016}. Virtually, the
offsets between the Ly$\alpha$ luminosity functions based on our two methods
are mostly within 1$\sigma$ uncertainties as seen from Figure \ref{f10.eps}.
The number densities based on the aperture-corrected fluxes and completeness
derived using stars are systematically low though. The main contributor of
the systematics comes from the completeness correction.  This is especially the
case for the faint part, as clearly shown in Figure
\ref{comparecompleteness_usestackimagemag_auto_usestarapercor.eps}.  Therefore,
we caution that systematics potentially introduced by different methods should
be considered when Ly$\alpha$ luminosity functions from different studies are
compared. Although number densities are affected by completeness corrections,
the slope of our Ly$\alpha$ luminosity functions at the faint end is consistent
with \citet{Konno+2016} and \citet{Sobral+2017}, regardless the use of measures
of total fluxes and completeness curves.

Regarding to the bright end ($L({\rm Ly\alpha}) > 10^{43}\, {\rm erg\,s^{-1}}$)  of
the luminosity functions, our results are consistent with \citet{Konno+2016}
but show higher volume densities than \citet{Sobral+2017}, as shown in both
Figure \ref{compareLFshapeliterature.eps} and Figure \ref{f10.eps}. The use of
MAG\_AUTO and completeness curves based on the reconstructed LAE image
does not change our bright-end luminosity functions dramatically.
\citet{Sobral+2017} claimed that the higher volume densities at the bright end
in \citet{Konno+2016} are caused by contaminations from non-LAE AGNs. Our AGNs
are spectroscopically confirmed $z\sim2.23$ LAEs, so our higher volume
densities than \citet{Sobral+2017} are real. But we note that the bright-end
luminosity functions suffer from small number statistics.  There are typically
1-2 objects in each bin. Virtually, the figures show that the bright-end
difference between our data points and the results in \citet{Sobral+2017} is
mostly within 1$\sigma$ errors. Furthermore, the cosmic variance uncertainties
are probably under-estimated since AGNs have much larger bias factor than our
LAEs \citep{Allevato+2011}. The actual error bars should be even larger, which
would lead to a conclusion that the bright end of our Ly$\alpha$ luminosity
functions is consistent with that of \citet{Sobral+2017} within 1$\sigma$
uncertainties.

\section{LAEs as Low Mass Galaxies at $z\sim2.23$}

The universe was in a critical epoch at $z\sim2$ when SFR
density reached the highest. Most studies at this redshift focused on massive
galaxies. Low mass galaxies are also important in constraining galaxy formation
models. Yet it is quite challenging to identify low-mass galaxies at high
redshifts. The narrowband technique provides an effective way to detect LAEs
with very low stellar masses at high redshifts. However, only a small fraction of them
are detected in longer wavelength bands \citep[e.g.,][]{Nilsson+2011}, making it
hard to estimate their stellar population parameters. We specifically chose
both COSMOS and ECDF-S fields where deep ancillary photometry data, especially
deep \textit{Spitzer}/IRAC observations, are available.  Therefore we are
able to study the properties of low-mass galaxies at $z\sim2$ by investigating
their stellar population, star formation property and dark matter halo mass.

\subsection{Stellar Population and Stellar Mass Function\label{subsec:SMF}}

\begin{figure}
\plotone{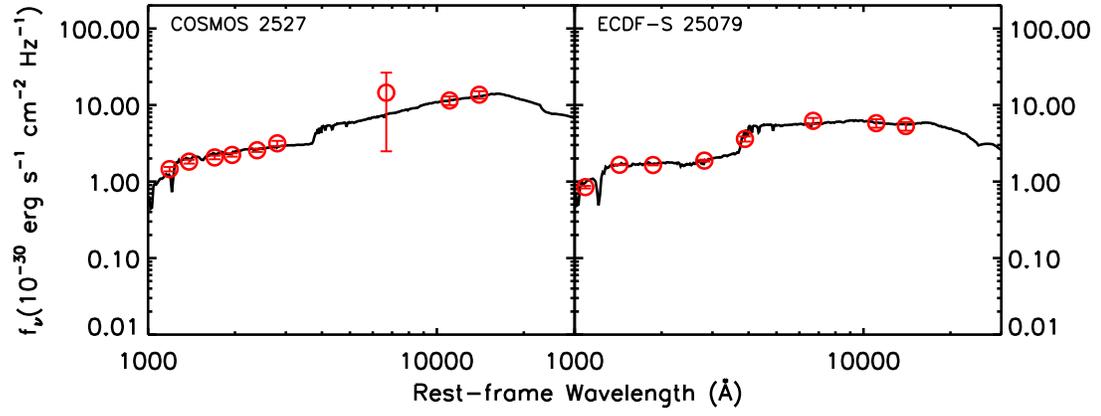}
\caption{Examples of SED fittings for IRAC-detected LAEs in COSMOS (left) and in ECDF-S (right). The red open circles are the observed flux densities
and the black curves are the best-fit SEDs. }
\label{LAE_withirac_plotsedfnu_example.eps}
\end{figure}

The \textit{Spitzer} Extended Deep Survey \citep[SEDS;][]{Ashby+2013} reaches
limiting flux densities of sub micro-Jansky in 3.6 \,$\mu$m and 4.5\,$\mu$m for
COSMOS and ECDF-S, providing critical constraints on stellar mass measurements
for faint LAEs in these two fields.  Out of the 446 sample non-AGN LAEs, 130
($\sim 29$\%) were detected in the SEDS imaging, which include 96 ($\sim 38$\%)
LAEs in the ECDF-S field and 34 ($\sim 18$\%) LAEs in the COSMOS field. The
different SEDS detection fractions in COSMOS and ECDF-S are due to the fact
that SEDS only covers a part ($10'\times60'$ strip) of our narrowband observed
area in COSMOS.  For the 130 IRAC-detected LAEs, we estimated their stellar
masses using the SED fitting software FAST \citep{Kriek+2009}.
Multi-wavelength SEDs were constructed as follows: The optical and NIR data are
retrieved from MUSYC \citep{Cardamone+2010}, GEMS \citep{Rix+2004}, Taiwan
ECDFS Near-Infrared Survey \citep[TENIS;][]{Hsieh+2012}, COSMOS
\citep{Capak+2007} and SEDS \citep{Ashby+2013}.  Specifically, MUSYC $U$, $B$
bands, GEMS F606W, F850LP bands, TENIS $J$, $K_{\rm s}$ bands and SEDS
3.6\,$\mu$m, 4.5\,$\mu$m are used for the ECDF-S field, while  $u^*$, $B_{\rm
J}$, $V_{\rm J}$, $r^+$, $i^+$, $z^+$, $J$, $K_{\rm s}$ bands from the COSMOS
photometry catalog and SEDS 3.6\,$\mu$m, 4.5\,$\mu$m are used for the COSMOS
field\footnote{For some LAEs, there are no photometric data available in a few
required wavebands.  But the remaining wavebands data, especially the crucial
MIR bands provide reasonable constraints on the SED shapes and hence the
stellar masses.}. The stellar population synthesis models of
\citet{Bruzual+2003} was adopted, assuming a Salpeter IMF, an exponentially
declining star formation history and a stellar metallicity of 0.2\,${\rm
Z_\odot}$. For dust attenuation modeling, the Calzetti dust extinction law
\citep{Calzetti+2000} was used. Figure
\ref{LAE_withirac_plotsedfnu_example.eps} shows examples of our SED fitting for
two IRAC-detected LAEs.  We note that H$\alpha$ emission contributes to the
$K_{\rm s}$ band flux densities that would potentially bias the stellar mass
estimates.  So we performed SED fitting without using $K_{\rm s}$ band data and found
that the stellar masses do not change much. In other words, no systematic
biases were introduced by including $K_{\rm s}$ band in the SED fitting. The
stellar masses were mainly constrained by the SEDS MIR data.

\begin{figure}
\plotone{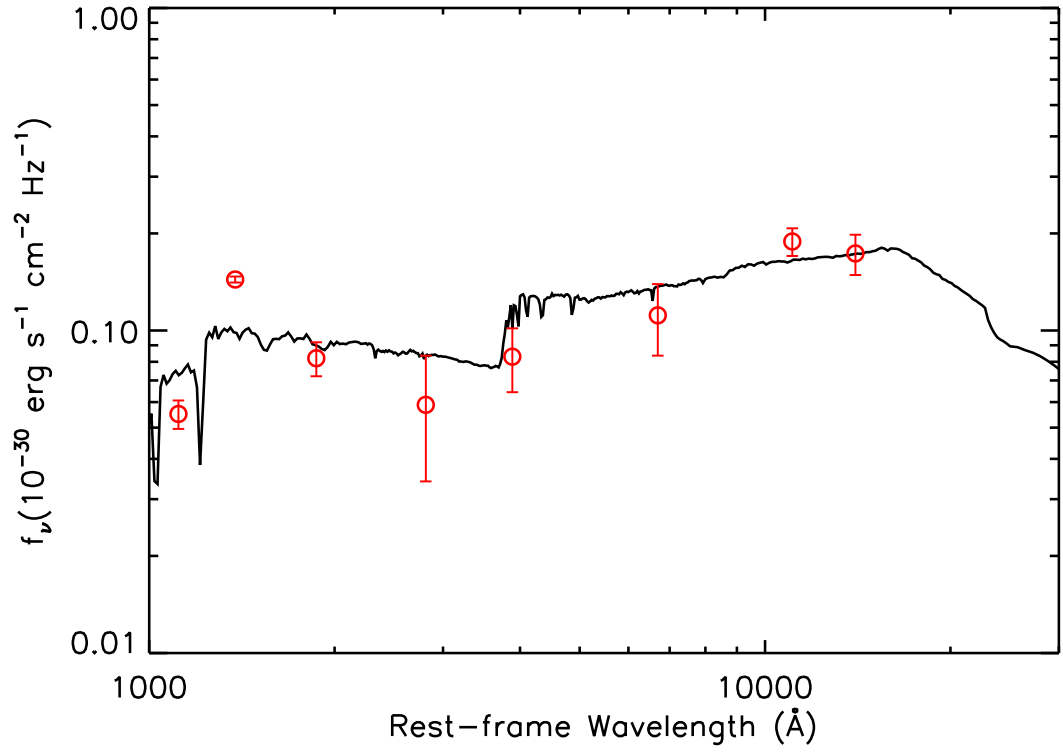}
\caption{SED fitting of the stacked result of IRAC-undetected LAEs. The red open circles are the stacked flux densities 
and the black curve represents the best-fit SED.}
\label{laestacking_ni_m_sedfitting_fnu.eps}
\end{figure}

Our SED fitting results show that the stellar masses for most IRAC-detected
LAEs are in the range of $8<{\rm log}(M_\star/M_\odot)<10$, but there do exist
massive LAEs with stellar mass larger than $10^{10}M_\odot$, even more massive
than $10^{11}M_\odot$.  On the other hand, for IRAC-undetected LAEs, we use
stacking analysis to derive a median SED. As mentioned above, SEDS only covers
a part of our narrowband imaging survey area for COSMOS. To achieve high
signal-to-noise ratio, we only used IRAC-undetected LAEs in ECDF-S for the
stacking analysis.  The stacked IRAC 3.6$\mu$m and 4.5$\mu$m magnitudes for the
IRAC-undetected LAEs are $26.93\pm0.09$ and $27.02\pm0.14$ mag, respectively,
which are similar to the stacking result for LAEs at $z\sim3.1$
\citep{Lai+2008}. Figure \ref{laestacking_ni_m_sedfitting_fnu.eps} presents the
median-stacked SED along with the best-fit, which yields a stellar mass of
${\rm log}(M_\star/M_\odot)=7.97^{+0.05}_{-0.07}$ and dust extinction of
$A_{\rm v}=0.12^{+0.25}_{-0.08}$ mag. This is among the lowest mean
stellar mass compared to the LAE samples in previous studies
\citep{Vargas+2014,Hagen+2014,Hagen+2016}. In Figure
\ref{laestacking_ni_m_sedfitting_fnu.eps}, we note that the $B$ band photometry
is far above the best-fit but we cannot figure out the reason for this
unreasonably high flux. Fortunately, the stellar mass and dust attenuation are
not affected by this photometric data point. If $B$ band photometry is not
included in the SED fitting, the results only change within 1$\sigma$ errors.

Although LAEs may just be one of the high-redshift galaxy populations with
${\rm log}(M_\star/M_\odot)\sim8$, there are few studies on galaxies with such
a low stellar mass at $z\sim2$ besides LAEs, thus poor constraints on the
low-mass end of the stellar mass function. With the estimation of a median
stellar mass of $\sim10^8M_\odot$ from the IRAC-undetected LAEs, we can have a
rough estimation of their number density. This was simply done by dividing
the number of IRAC-undetected LAEs by the product of the effective co-moving
volume and the stellar mass range, without any corrections. Since stellar mass
cannot be derived for IRAC-undetected LAEs individually, the stellar mass range
cannot be estimated in the normal way. Instead, we estimated the stellar mass
range via the star formation rate range. The stellar mass range in log-space
is equal to the logarithmic star formation rate range under the assumption that
the specific SFRs (i.e., $SFR/M_\star$) of these IRAC-undetected LAEs are the same. We
estimated the star formation rate range covered by the IRAC-undetected LAEs
using the rest-frame UV-derived SFRs (c.f. Section \ref{subsec:SF}) of these
LAEs and obtained the number density of $\rm {log}(\Phi/{\rm Mpc^{-3}
dex^{-1}})=-3.0$ at ${\rm log}(M_\star$/$M_\odot)=8$.  We note that the number
density derived this way suffers from large uncertainties introduced by the
sample incompleteness and the assumption made in deriving the stellar mass
range. It is probably a lower limit of the real number density considering the
sample incompleteness. But it is still worth examining its position on the
stellar mass function diagram given its low stellar mass.

\begin{figure}
\plotone{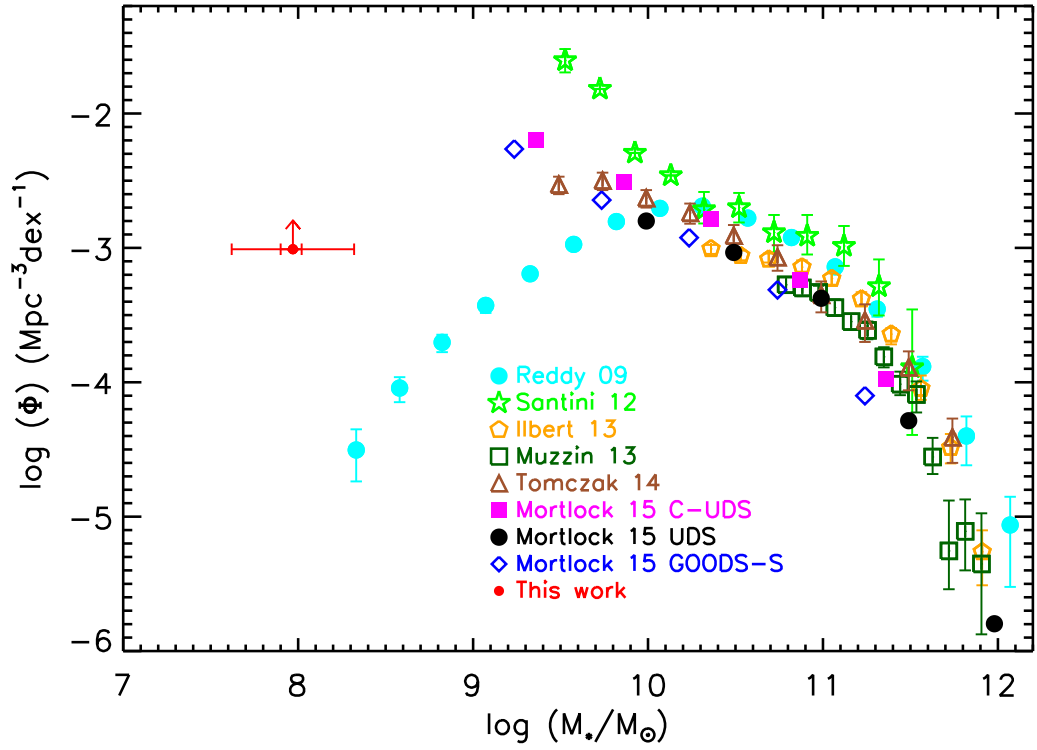}
\caption{Stellar mass functions at $z\sim2$. The red symbol represents the constraints from our IRAC-undetected LAEs.
The smaller error bars indicate the $1\sigma$ errors from the SED fitting, while the larger error bars reflect the possible
rms scatter in the stellar mass distribution of the IRAC-undetected LAEs, derived from the rms scatter in the SFRs of these LAEs
by assuming that their specific SFRs are the same.
Deep stellar mass functions from recent studies as indicated at the lower-right corner are shown for comparison.}
\label{compareSMF.eps}
\end{figure}

Figure \ref{compareSMF.eps} shows the derived number density of our
IRAC-undetected LAEs in the stellar mass function diagram, in comparison with
the existing stellar mass functions at redshift $z\sim2$ \citep{Reddy+2009,
Santini+2012, Ilbert+2013, Muzzin+2013, Tomczak+2014, Mortlock+2015}.  As can
be seen from Figure \ref{compareSMF.eps}, the only number density measurement
to ${\rm log}(M_\star$/$M_\odot)=8$ are based on the BX redshift sample
\citep{Reddy+2009} and are much lower than ours.  All photometric redshift
samples at $z\sim2$ are not deep enough to reach this mass limit.  However,
these stellar mass functions at ${\rm log}(M_\star$/$M_\odot)=9.5$ are already
a factor of $\sim4-5$ higher than our LAE at ${\rm log}(M_\star$/$M_\odot)=8$.
Therefore, our estimation provides a lower limit for the mass function at this
mass limit.

If the extrapolation of the previous stellar mass functions to the low-mass end
traces the real stellar mass function, we must answer the question: which types
of galaxies compose of the low-mass galaxy population? Besides the LAEs,
possible populations are galaxies with large amount of dust, galaxies that have
been quenched or genuine star-forming galaxies with dust/gas geometries and
orientations preventing Ly$\alpha$ photons escaping from the galaxies.
However, dusty galaxies tend to be massive \citep[e.g.,][]{Whitaker+2012,
Whitaker+2014}, and thus it is impossible for them to be a major population at
the low-mass end of the stellar mass function.  On the other hand, it is also
unlikely that the quenched galaxies dominate the low-mass end of stellar mass
function \citep[e.g.,][]{Tomczak+2014}. The most possible missing low-mass
galaxies are star-forming galaxies with interstellar medium (ISM) geometries
and orientations against the escape of Ly$\alpha$ photons. It is difficult to
identify the fraction of such galaxies at the moment because of their faint
continua.  Future deep observations in NIR/MIR by the James Webb Space
Telescope (JWST) may shed light on this.

\subsection{Star Formation Properties\label{subsec:SF}}

The SFMS relation and its extrapolation towards low-mass end have been widely
used to characterize the star formation mode of LAEs.  However, agreement has
not been reached on the locations of LAEs relative to the SFMS relation
\citep{Guaita+2013, Hagen+2014,Vargas+2014, Oteo+2015, Hagen+2016,
Shimakawa+2017, Kusakabe+2018}.  Differences in the narrowband survey depths
and the adoption of extinction curves may be responsible for the discrepancy
\citep{Shimakawa+2017, Kusakabe+2018}.  Given that our survey is among the
deepest narrowband surveys and our sample is the largest LAEs sample with
individual stellar mass measurements, it is worth revisiting the SFMS relation
based on our LAEs sample.

The rest-frame FUV luminosity is the most commonly used SFR tracer at high
redshifts. We calculated the rest-frame FUV luminosities for our sample LAEs
from the observed $B$-band flux densities. Then the FUV luminosities were
converted to SFRs using a conversion factor derived using STARBURST99 version
7.0.1 \citep{Leitherer+1999, Leitherer+2010, Leitherer+2014, Vazquez+2005}
synthesis models for a constant star-forming population with age of 100\,Myr
and 0.2\,$Z_\odot$ stellar metallicity: \begin{equation} {\rm SFR}({\rm
M}_\odot {\rm \,yr}^{-1})=1.35\times10^{-28} L_{\nu} ({\rm erg\,s}^{-1}{\rm
\,Hz}^{-1}).  \end{equation} Dust attenuations were not accounted for in the
SFRs derived here.  The dust-uncorrected SFRs for IRAC-detected LAEs in both
COSMOS and ECDF-S are in a range of ${\rm 0.1<SFR<10}$ M$_\odot$\,yr$^{-1}$,
while for the LAEs without IRAC detections, the median SFR is
${\rm log} (SFR/M_\odot$\,yr$^{-1})=-0.14$ with a rms scatter of 0.35 dex. In
addition, six massive LAEs with stellar masses about $10^{11} M_{\odot}$ in the
two fields are detected at MIPS 24\,$\mu$m  or even at \textit{Herschel} FIR
bands.  We calculated their SFRs using the MIPS 24\,$\mu$m flux densities
\citep{Rieke+2009}, which are so high that they are qualified as Luminous
Infrared Galaxies (LIRGs) or even Ultra-Luminous Infrared Galaxies (ULIRGs).
Considering the large PSFs of the MIR/FIR images, we inspected the
\textit{Spitzer}/IRAC 3.6\,$\mu$m and MIPS 24\,$\mu$m images for these six LAEs
and confirmed that their high MIR emissions are not from contaminations by
neighboring bright objects.

\begin{figure}
\includegraphics[width=0.50\textwidth]{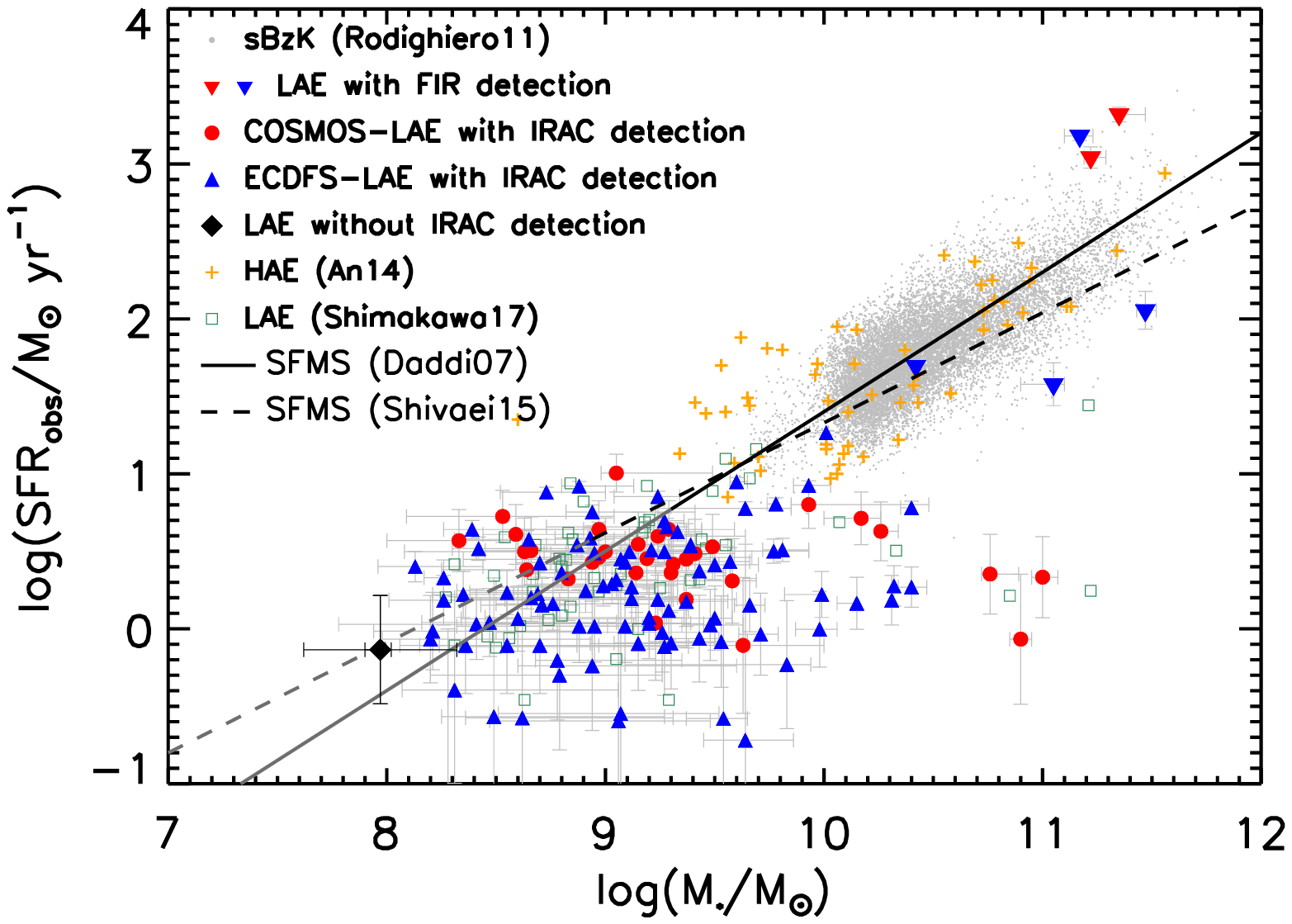}
\hfil
\includegraphics[width=0.50\textwidth]{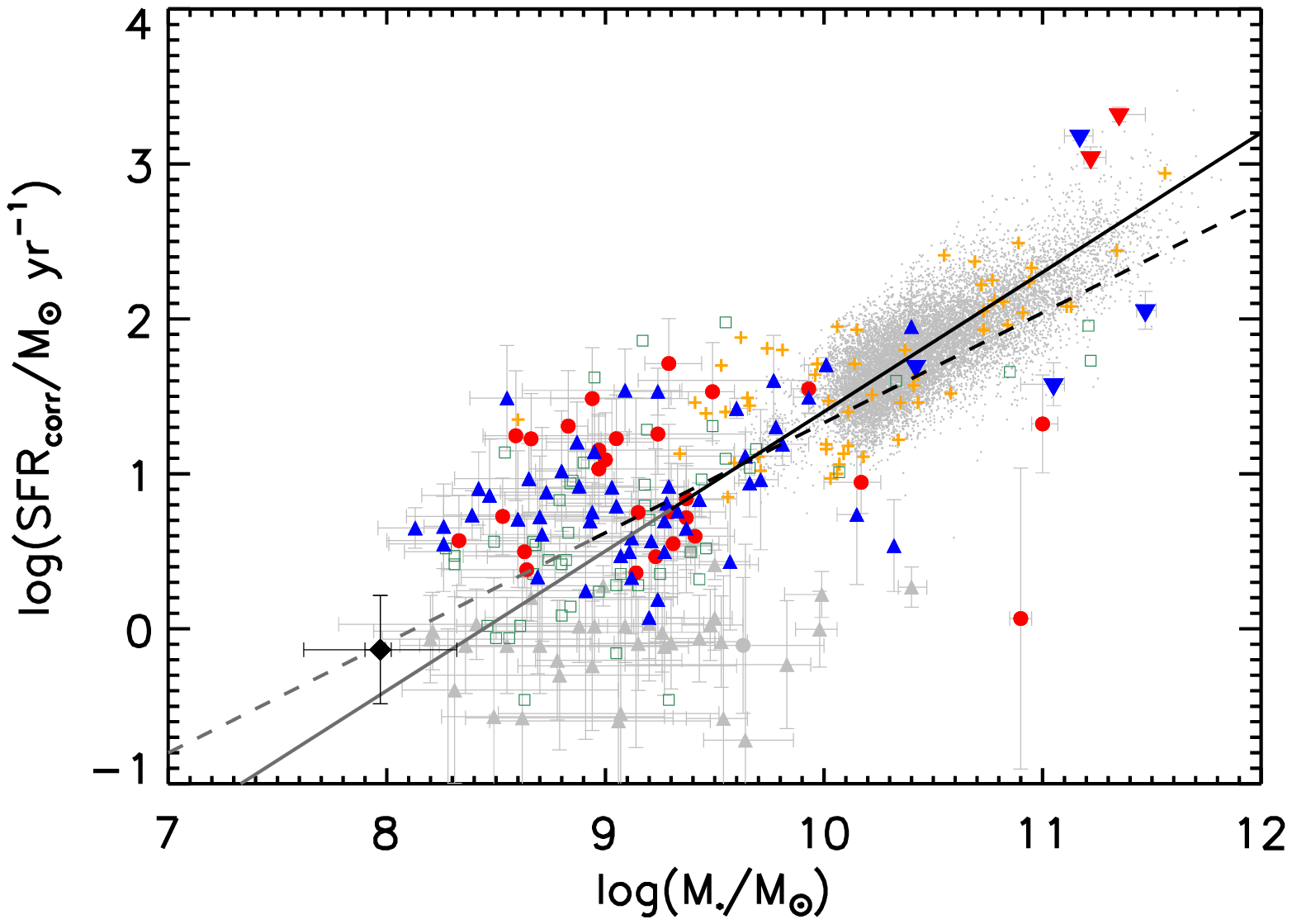}
\caption{SFR versus stellar mass relation for our IRAC-detected LAEs (red filled circles for COSMOS and blue filled triangles for 
ECDF-S) and the stacked results (black diamond) for IRAC-undetected LAEs. For the data point representing the stacked results (the black diamond), 
the horizontal error bars have the same meaning as those in Figure \ref{compareSMF.eps}, while the vertical error bars
indicate the rms scatter in the SFR distribution of the IRAC-undetected LAEs.
LAEs at $z\sim2.5$ from \citet[][dark green open squares]{Shimakawa+2017}
are plotted for comparison. BzK-selected star-forming
galaxies \citep[gray dots;][]{Rodighiero+2011}  and H$\alpha$ emitters in the ECDF-S field \citep[orange crosses;][]{An+2014} 
are also plotted. Our LAEs with MIR/FIR detections are represented by red (for COSMOS) and blue (for ECDF-S) upside-down triangles. 
The star formation main sequence relations
at $z\sim2$ from \citet{Daddi+2007} and \citet{Shivaei+2015} are plotted as the black solid and dashed lines, respectively.
Their extrapolations towards the low-mass regime are plotted using respective lines in gray.
{\it left:} The SFRs for both LAEs in this work
and those in \citet{Shimakawa+2017} were not corrected for dust attenuations. SFRs of the others were dust corrected as in the
original paper. {\it right:} For IRAC-detected LAEs with $B$, $V$, $R$, $I$ photometry, the SFRs were corrected for dust attenuation 
using the UV slope and Calzetti law \citep{Calzetti+2000}. For IRAC-detected LAEs without $B$, $V$, $R$, $I$ photometry, no dust attenuation was performed and
they are plotted as gray symbols. No dust correction was done for the stacked result of the IRAC-undetected LAEs. All the SFRs from the literature 
were dust-corrected as in the original paper.}
\label{LAE_newmass_SFR_final_witherrorbar.eps}
\end{figure}

The left panel of Figure \ref{LAE_newmass_SFR_final_witherrorbar.eps} shows the
locations of our sample LAEs on the dust-uncorrected SFR versus stellar mass
diagram.  The red solid circles and blue solid triangles represent LAEs in the
COSMOS and ECDF-S fields, respectively, while the large red and blue inverted
triangles denote the MIR/FIR-detected LAEs in the respective fields.  For
comparison, we also plot $z\sim2$ BzK-selected galaxies with dust-corrected
SFRs \citep{Rodighiero+2011}, H$\alpha$ emitters with dust-corrected SFRs
\citep[HAEs;][]{An+2014} and the 50 LAEs in \citet{Shimakawa+2017} with
dust-uncorrected SFRs.  The widely used $z\sim2$ SFMS relations from \citet{Daddi+2007}
and \citet{Shivaei+2015} are overplotted in this figure.  It is
clear from the left panel of Figure
\ref{LAE_newmass_SFR_final_witherrorbar.eps} that the majority of our
IRAC-detected LAEs and the stacked IRAC-undetected LAEs are located on the SFMS
relations or their extrapolations towards low stellar mass end, within
1$\sigma$ scatters (0.3 dex).  In addition, the six MIR/FIR-detected massive
LAEs with high SFRs are also on the SFMS relation and mix with the BzK-selected
galaxies. The left panel of Figure \ref{LAE_newmass_SFR_final_witherrorbar.eps}
also shows that our 130 IRAC-detected LAEs are well mixed with the 50 LAEs of
\citet{Shimakawa+2017}.  Nevertheless, we note that as the stellar mass
increases, the fraction of LAEs below SFMS also increases.  Especially, as the
stellar mass is larger than a few times $10^{9} M_{\odot}$, almost all LAEs are
below the SFMS line. It may imply that more massive LAEs tend to be dustier and
dust attenuation corrections are necessary for LAEs. 

\begin{figure}
\plotone{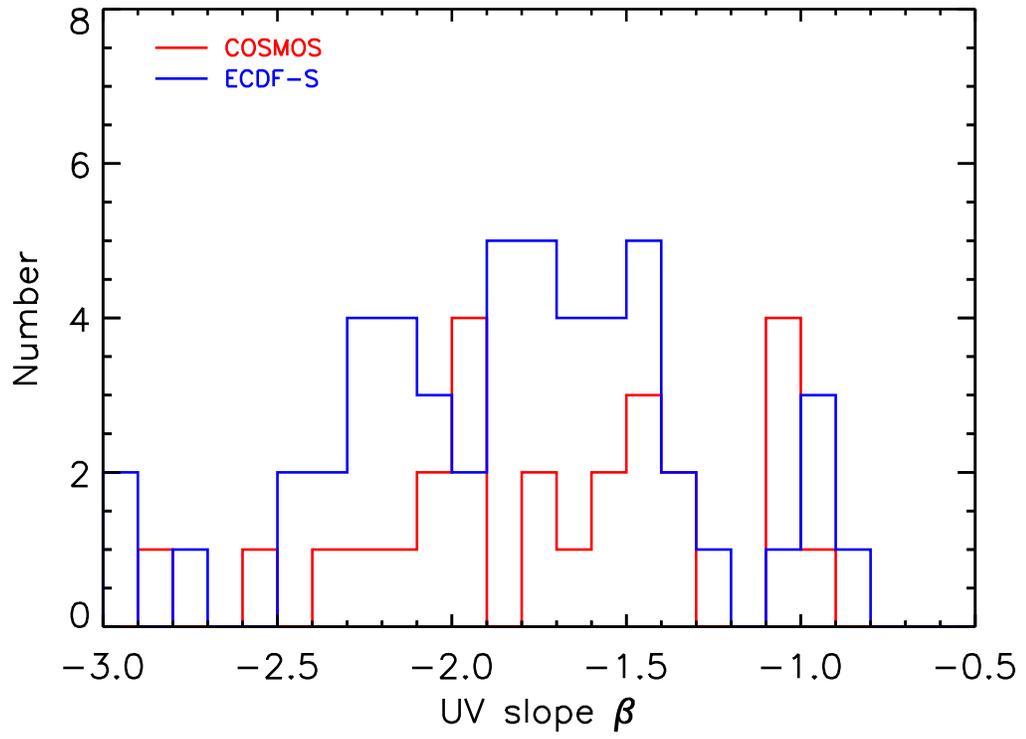}
\caption{Distribution of the UV slope $\beta$ for IRAC-detected LAEs with $B$, $V$, $R$ and $I$ measurements for the COSMOS (red) and ECDF-S (blue) fields.}
\label{compare_UVbetadis.eps}
\end{figure}

Currently, dust attenuations for LAEs are derived from SED fitting or from the
UV slope $\beta$ ($f_\lambda \varpropto \lambda^\beta $).  Given that the color
excess E(B-V) derived via SED fitting is affected by the stellar mass
\citep{Shivaei+2015}, we estimate E(B-V) from the UV slope $\beta$ in this
work.  The UV slope $\beta$ could be determined reasonably well if several
wavebands data covering the rest-frame 1300--2600\,\AA\ are available. This
wavelength range corresponds to the $B$, $V$, $R$ and $I$ bands for objects at
$z\sim2.23$. For the 34 IRAC-detected LAEs in the COSMOS field, 33 are in the
COSMOS public catalog and have deep $B_{\rm J}$, $V_{\rm J}$, $r^+$, $i^+$
bands measurements, while for the ECDF-S field, only 58 out of 96 IRAC-detected
LAEs have MUSYC $B$, $V$, $R$ and $I$ photometry. The six MIR/FIR-detected LAEs
are included in the subsample with $B$, $V$, $R$ and $I$ bands photometry.  We
obtained $\beta$ by fitting a power-law to the four bands flux densities via
chi-square minimization\footnote{For four objects in COSMOS and three objects
in ECDF-S, their UV continuum are suspicious and the four wavebands power-law
fitting is with large errors in the best-fit values, i.e.  larger than 50\%.
Thus these objects were excluded in the dust-correction related analysis.}.
After excluding the six MIR/FIR-detected LAEs and LAEs with unreliable $\beta$
measurements, we were left with 27 LAEs in COSMOS and 51 LAEs in ECDF-S with
reliable $\beta$. The uncertainties in $\beta$ vary from 2\% to 49\% and have
been incorporated into the total error budget in dust-corrected SFRs
subsequently. The Calzetti extinction law \citep{Calzetti+2000} was then used
to calculate E(B-V) from $\beta$ under the assumption that the intrinsic UV
slope $\beta_0$ is -2.23 \citep{Meurer+1999}. For objects with $\beta < -2.23$,
zero dust extinctions were assumed. Figure \ref{compare_UVbetadis.eps} shows
the distributions of $\beta$ with a median of $-1.8$ for the 78 IRAC-detected
LAEs with $B$, $V$, $R$ and $I$ measurements.  As can be seen, there are 15
LAEs with $\beta < -2.23$ and hence zero dust extinctions.  Figure
\ref{compare_UVbetadis.eps} also reveals a broad distribution in $\beta$ for
LAEs in both fields, which leads to a wide range of E(B-V) varying from 0 to
0.3 mag, with a median value of 0.1 mag. Using the E(B-V), we obtained the
dust-corrected SFRs for the 78 IRAC-detected LAEs spanning a range of ${\rm
1<SFR<100}$\,$M_\odot$\,yr$^{-1}$. It should be noted that due to the lack of
$B$, $V$, $R$ and $I$ photometry, dust corrections were not performed for 39
IRAC-detected LAEs. The 39 LAEs have lower dust-uncorrected SFRs than most of
the LAEs that have $B$, $V$, $R$ and $I$ measurements.

The right panel of Figure \ref{LAE_newmass_SFR_final_witherrorbar.eps} presents
the 84 LAEs with dust-corrected SFRs, including the six MIR/FIR-detected LAEs
whose SFRs were calculated from 24\,$\mu$m fluxes and the 15 LAEs with zero
dust extinctions. The LAEs in \citet{Shimakawa+2017} are also plotted with
dust-corrected SFRs.  The symbols are the same as those in the left panel of
Figure \ref{LAE_newmass_SFR_final_witherrorbar.eps}.  Note that we also plot
the 39 LAEs without dust corrections for their SFRs in the figure by solid gray
circles or triangles.  As shown in the right panel of Figure
\ref{LAE_newmass_SFR_final_witherrorbar.eps}, most LAEs with dust-corrected
SFRs are located along the SFMS within 1$\sigma$ scatter, although a small
fraction of LAEs are located above the SFMS, indicating that they are in active
star formation mode.  As for the stacked IRAC-undetected LAEs, we do not
perform dust corrections because they have minor dust attenuation as derived
from the SED fitting.  It is obvious that these low-mass LAEs sit on the
low-mass extrapolations of the SFMS.  On the other hand, the LAEs with stellar
mass larger than $10^{10} M_{\odot}$, along with the MIR/FIR-detected
(U)LIRGs-like LAEs, are on the SFMS as well.  In summary, LAEs are
heterogeneous populations that have stellar masses and SFRs covering more than
three orders of magnitude, i.e., ${\rm 8 < log(M_\star/M_\odot) < 11.5}$, ${\rm
1<SFR<2000}$\,M$_\odot$\,yr$^{-1}$ and suffer from dust extinctions spanning a
wide range.  However, LAEs are mostly low-mass star-forming galaxies and they
follow the SFMS relations defined by massive normal star-forming galaxies and
their extrapolations to the low mass regime. This suggests that they are normal
star-forming galaxies, instead of a special galaxy population in terms of star
formation modes. It is unusual that an LAE is massive and MIR/FIR luminous,
since even a small amount of dust could stop Ly$\alpha$ photons from escaping
the galaxy. Such dusty, massive LAEs may have special dust/gas geometries
favoring the escape of Ly$\alpha$ photons, as suggested by studies on
Ly$\alpha$ and optical emission line profiles from local ULIRGs
\citep{Martin+2015}.

In the literature, different conclusions have been drawn on the relations of
LAEs with respect to the SFMS. Survey depths and use of extinction curves have
been proposed to be the causes. Since \citet{Shimakawa+2017} have comparable
narrowband survey depth with ours, we overplotted their LAEs with and without
dust attenuation corrections in SFRs in the left and right panels of Figure
\ref{LAE_newmass_SFR_final_witherrorbar.eps} for comparison. We can see from
the left panel of Figure \ref{LAE_newmass_SFR_final_witherrorbar.eps} that the
two LAE samples cover almost the same range in both the stellar mass and the
dust-uncorrected SFR.  Virtually, the two sample LAEs are mixed together in the
dust-uncorrected SFR versus stellar mass diagram. On the other hand, the right
panel of Figure \ref{LAE_newmass_SFR_final_witherrorbar.eps} shows that the
LAEs from the two samples are mostly mixed well except a lack of our LAEs in
the low SFRs part, which is caused by the absence of broad $B$, $V$, $R$ and
$I$ bands photometry for low SFR objects. Furthermore, we inspected Figure 10
of \citet{Hagen+2014} who studied LAEs with high Ly$\alpha$ luminosities
(L(Ly$\alpha$) $> 10^{43}$\,erg\,s$^{-1}$), at which there are almost no LAEs
below the SFMS line.  In comparison with the right panel of Figure
\ref{LAE_newmass_SFR_final_witherrorbar.eps} in this work, the absence of LAEs
below the SFMS in \citet{Hagen+2014} seems to be caused by the selection
effect that relatively shallow narrowband surveys leave out galaxies with lower
SFRs, as pointed out by \citet{Oyarzun+2017}. Regarding to the adoption of
extinction laws, both \citet{Shimakawa+2017} and this work use the Calzetti
extinction curve and find that the LAEs are not significantly above the SFMS
relations in the dust-corrected SFR versus stellar mass diagram. Therefore, it
seems that it is not a necessity to employ a different extinction law.

\subsection{Dark Matter Halo Mass\label{subsec:clustering}}

\begin{figure}
\plotone{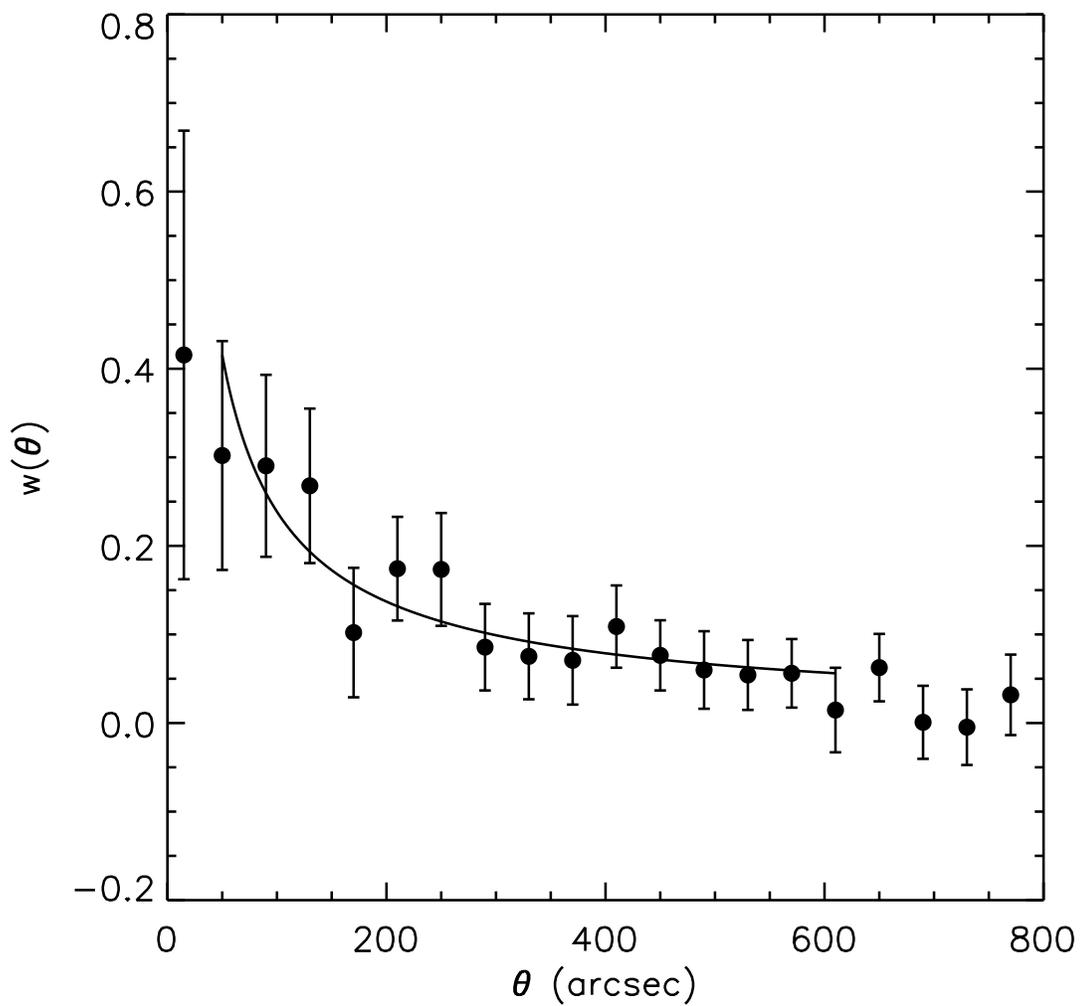}
\caption{Angular two-point correlation function of the whole sample.
The curve is the best power-law fit.}
\label{atpcf_ham.eps}
\end{figure}

In the $\Lambda$CDM paradigm, galaxies form in dark matter halos,
and galaxy evolution is closely linked to its hosting dark matter
halo mass. In this subsection, we derive the dark matter halo mass for our
LAEs. The bias factor and dark matter halo mass of our LAEs sample were estimated via
clustering analysis following \citet{Guaita+2010} and \citet{Kusakabe+2018}.
First, we calculated the angular two-point correlation function using the
Landy-Szalay estimator \citep{Landy+1993}: 
\begin{equation}
w (\theta) = \frac{DD(\theta)-2DR(\theta)+RR(\theta)}{{RR(\theta)}},
\end{equation} 
where DD, DR and RR are the normalized counts for data-data,
data-random, and random-random pairs, respectively. We generated a random sample
that is 200 times the LAE sample size with the same geometry.  A power law form
$w(\theta) = A \theta^{-\beta}$ was assumed for the angular correlation
function. However, due to the limited size of the survey area, the observed
angular correlation function is actually $w(\theta)-AC=A(\theta^{-\beta}-C)$,
where AC is the integral constraint. By performing a Monte Carlo integration,
we can first estimate C \citep[e.g.,][]{Roche+1999} and then fit
$A(\theta^{-\beta}-C)$ to the data. The clustering amplitude A was thus obtained from
the fitting by further fixing $\beta$ to 0.8 following the literature \citep[e.g.,][]{Guaita+2010, 
Matsuoka+2011, Coupon+2012}. Note that we
only used a selected range of $\theta$ ($50\arcsec \lesssim  \theta \lesssim 600\arcsec$)
during the fitting, in order to avoid the influence of the one-halo term at
small scales ($\theta < 50\arcsec$) and sampling noise at large scales.
The best-fit values of A and the integral constraint are $9.5\pm2.2\,arcsec^{0.8}$ and $0.06$,
respectively.  The angular two-point correlation function, along with the
best-fit curve of our LAEs are shown in Figure \ref{atpcf_ham.eps}.

Corresponding to the power law form of $w(\theta)$, the spatial correlation
function has the form of $\xi(r)=(r/r_0)^{-(\beta+1)}$. Assuming a Gaussian
distribution of the LAE redshifts within our narrowband window, we obtained the
real space correlation length according to \citet{Simon2007}, which is $3.66\pm0.47$
Mpc. Then we calculated the bias factor of LAEs by
$b=\sqrt\frac{\xi(r)}{\xi_{\rm DM}(r)}$, where $\xi(r)$ and $\xi_{\rm DM}(r)$ are the
correlation function of LAEs and the underlying dark matter in the linear
theory, respectively. Here r is chosen to be $8h^{-1}$ Mpc, following \citet{Ouchi+2003}
and \citet{Kusakabe+2018}.  The resultant bias factor is
$1.31\pm0.15$.  Finally, the halo mass $M_{\rm h}$ was obtained via the relation
between bias factor and the peak height in the linear density field
$\nu=\delta_c/ \sigma(M_{\rm h},z)$ \citep{Tinker+2010}, where $\delta_c=1.686$ is
the critical overdensity for dark matter collapse and $\sigma(M_{\rm h},z)$ is the
rms fluctuation in a sphere that encloses mass $M_{\rm h}$ on average at present
time, extrapolated to redshift $z$ with the linear theory. The bias factor
derived above corresponds to a mean dark matter halo mass of $\log(M_{\rm
h}/M_{\odot}) = 10.8^{+0.26}_{-0.42}$.  Note that the errors reported here do
not account for cosmic variance.  Since our survey area (COSMOS and ECDF-S
fields) is just $\sim0.34$ ${\rm deg}^2$,  cosmic variance should be important,
as discussed by \citet{Kusakabe+2018}. According to the scaling relation in
\citet{Kusakabe+2018}, we estimated an uncertainty of $\sim46\%$ due to cosmic
variance in the bias factor, resulting in a bias factor of $1.31\pm0.34$ and
halo mass of $\log(M_{\rm h}/M_{\odot}) =10.8^{+0.56}_{-1.1}$.  Most recently,
based on a large sample (1937 LAEs) of $z\sim2.2$ LAEs with NB387$_{\rm
tot}\leq25.5$ mag in four survey fields covering a total area of $\simeq1$
${\rm deg}^2$, \citet{Kusakabe+2018} obtained a bias factor of
$1.22^{+0.23}_{-0.26}$ and halo mass of $\log(M_{\rm h}/M_{\odot})
=10.6^{+0.5}_{-0.9}$. Accordingly, they predicted that in the local universe
their LAEs would be typically hosted by dark matter halos with mass comparable
to that of the Large Magellanic Cloud (LMC).  The bias factor and halo mass
based on our 446 LAEs are consistent with those of \citet{Kusakabe+2018}
within $1\sigma$, although the errors in our analysis are larger due to the
smaller survey area. Therefore, the dark matter halo hosting our LAEs may
similarly evolve into a LMC-like halo at $z=0$.

\section{SUMMARY} 

We have conducted deep narrowband surveys for the COSMOS and ECDF-S fields to
search for Ly$\alpha$ emitters (LAEs) at redshift $z=2.23\pm0.03$ using our
customized narrowband filter $N3928$\,\AA\ at Megacam/Magellan II telescope.
Our observations reached a 5$\sigma$ limiting magnitudes in a 3\arcsec\
diameter aperture of $\sim26$ mag and a seeing FWHM of 0\farcs6.  Using
archival broad $U$ and $B$ bands images as a measure of the underlying
continuum, we selected 194 (including 4 AGNs) and 258 (including 2 AGNs) LAEs
over the 602 arcmin$^2$ and 613 arcmin$^2$ survey areas on the COSMOS and
ECDF-S fields, respectively. Our LAEs sample provides reliable measurements of
the Ly$\alpha$ luminosity function over the Ly$\alpha$ luminosity range of
$10^{41.8}-10^{42.8}$ erg\,s$^{-1}$. Within this luminosity range, the
Ly$\alpha$ luminosity functions of the COSMOS and ECDF-S fields are in a good
agreement with each other. The overall shapes of our Ly$\alpha$ luminosity
functions are consistent with that of \citet{Konno+2016} and
\citet{Sobral+2017} based on larger area ($1.43\,{\rm deg^2}$) Ly$\alpha$
surveys at similar redshifts. Thus our Ly$\alpha$ luminosity functions lend
further support to the steep faint-end slope. 

The existing multi-wavelength data from the rest-frame UV to the IR, especially
the deep \textit{Spitzer}/IRAC MIR data, allow us to explore the stellar
populations and star formation properties of LAEs. The \textit{Spitzer}
Extended Deep Survey (SEDS) provides important constraints on the stellar mass
estimates. For 29\% of our LAEs that were detected by IRAC at 3.6 \,$\mu$m or
4.5 \,$\mu$m, their stellar masses are in the range of $8 <{\rm
log(M_\star/M_\odot)}< 11.5$.  On the other hand, the SED fitting to the stacked
SED of the IRAC-undetected LAEs indicates a stellar mass of ${\rm
log}(M_\star/M_\odot)=7.97^{+0.05}_{-0.07}$ and dust extinction of $A_{\rm
v}=0.12^{+0.25}_{-0.08}$ mag. Based on the measurement of the median stellar
mass for the IRAC-undetected LAEs, we roughly estimate their mean number
density as $\rm {log}(\Phi/{\rm Mpc^{-3}
dex^{-1}})=-3.0$ at log($M_\star$/$M_\odot$)=8.  Although it is a
lower limit and much smaller than the extrapolation of the existing stellar
mass functions, it serves as an important observational constraint at such
low-mass regime.

Rest-frame FUV luminosities calculated from the observed $B$-band flux
densities were used to derive SFRs. The dust attenuations were estimated from
the UV slope $\beta$, based on public $B$, $V$, $R$ and $I$ bands photometry.
The dust-corrected SFRs of our LAEs cover a range of ${\rm
1<SFR<100}$\,M$_\odot$\,yr$^{-1}$, with six \textit{Spitzer}/MIPS 24\,$\mu$m or
even \textit{Herschel} FIR detected LAEs having SFRs up to 2000\,M$_\odot$\,yr$^{-1}$.
Although LAEs are heterogeneous populations that have stellar mass and SFR
covering more than three orders of magnitude, i.e.  ${\rm 8 <
log(M_\star/M_\odot) < 11.5}$, ${\rm 1<SFR<2000}$\,M$_\odot$\,yr$^{-1}$, they
are mostly composed of low-mass galaxies and follow the star formation main
sequence relations and their extrapolations to the low mass end. This indicates
that the star formation in most LAEs is taking place in a steady mode.

The two-point correlation function analysis for our LAEs sample yields a bias
factor of $1.31\pm0.34$ and corresponding dark matter halo mass of
$\log(M_{h}/M_{\odot}) =10.8^{+0.56}_{-1.1}$, which is consistent with those of
\citet{Kusakabe+2018} based on a much larger sample and survey area.  

\acknowledgments We would like to thank the anonymous referee for very helpful
comments and suggestions that improved the paper. We also thank Drs. Haruka
Kusakabe, Hong Guo, Jun Pan, Jie Wang for helpful discussions.  We acknowledge
Dr. Giulia Rodighiero for kindly providing their data for BzK samples in the
star formation main sequence diagram and Dr. Maureen Conroy for instructing us
to run the TCL script ``megawcs''.  This work is supported by the National Key
Research and Development Program of China (No. 2017YFA0402703) and the National
Natural Science Foundation of China (NSFC, No. 11373027 and 11733002). X.Z.Z.
thanks support from the NSFC (No. 11773076) and the Chinese Academy of Sciences
(CAS) through a grant to the CAS South America Center for Astronomy (CASSACA)
in Santiago, Chile. C.J. acknowledges support from the NSFC (No. 11773051),
and the CAS Key Research Program of Frontier Sciences (No. QYZDB-SSW-SYS033).
C.L. acknowledges the support by National Key Basic Research Programs of China
(No. 2015CB857004) and National Key R\&D Program of China (No. 2018YFA0404502),
and the NSFC (No.  11173045, 11233005, 11325314, 11320101002). This research
uses data obtained partially through the Telescope Access Program (TAP), which
has been funded by the National Astronomical Observatories of China, the
Chinese Academy of Sciences, and the Special Fund for Astronomy from the
Ministry of Finance.

\clearpage

\pagebreak

\begin{table}
\begin{center}
\caption{Narrowband observation parameters}
\begin{tabular}{lcccc}
\hline
Field & RA\tablenotemark{a} & DEC\tablenotemark{a} & Total exposure (min) & PSF(FWHM) \\
\hline
ECDF-S & 3$^{\rm h}$32$^{\rm m}$26.0$^{\rm s}$ & -27$\degr$49\arcmin20\arcsec & 660 & 0.6\arcsec \\
COSMOS & 10$^{\rm h}$00$^{\rm m}$27.9$^{\rm s}$ & +2$\degr$12\arcmin 03\arcsec & 600 & 0.6\arcsec \\
\hline
\end{tabular}
\end{center}
\tablenotetext{a}{This indicates the center of the pointing.}
\end{table}

\begin{table}
\begin{center}
\caption{Number counts of LAEs}
\begin{tabular}{lcc}
\hline\noalign{\smallskip}
m(Ly$\alpha$) & dN/dm  & (dN/dm)$_{\rm corr}$ \\
(mag)  &  ($10^{-2}$\,mag$^{-1}$\,arcmin$^{-2}$) & ($10^{-2}$\,mag$^{-1}$\,arcmin$^{-2}$) \\
\noalign{\smallskip}
\hline\noalign{\smallskip}
&  COSMOS & \\
\noalign{\smallskip}\hline
\noalign{\smallskip}
   22.25 &   0.33$^{+0.76}_{-0.27}$ &   0.33$^{+0.76}_{-0.27}$  \\
   22.75 &   1.00$^{+0.97}_{-0.54}$ &   1.00$^{+0.97}_{-0.54}$  \\
   23.25 &   0.33$^{+0.76}_{-0.27}$ &   0.33$^{+0.76}_{-0.27}$  \\
   23.75 &   1.99$^{+1.19}_{-0.79}$ &   1.99$^{+1.19}_{-0.79}$  \\
   24.25 &   1.99$^{+1.19}_{-0.79}$ &   1.99$^{+1.19}_{-0.79}$   \\
   24.75 &   8.97$^{+2.08}_{-1.71}$ &   8.97$^{+2.08}_{-1.71}$  \\
   25.25 &   13.96$^{+2.50}_{-2.14}$ &   13.97$^{+2.50}_{-2.14}$  \\
   25.75 &   18.27$^{+2.81}_{-2.46}$ &   19.45$^{+2.99}_{-2.62}$  \\
   26.25 &   14.95$^{+2.58}_{-2.22}$ &   22.50$^{+3.88}_{-3.34}$  \\
\noalign{\smallskip}
\hline\noalign{\smallskip}
&  ECDF-S & \\
\noalign{\smallskip}\hline
\noalign{\smallskip}
   22.25 &   ... & ... \\
   22.75 &   0.33$^{+0.75}_{-0.27}$ &  0.33$^{+0.75}_{-0.27}$ \\
   23.25 &   0.65$^{+0.86}_{-0.42}$ &  0.65$^{+0.86}_{-0.42}$ \\
   23.75 &   1.31$^{+1.03}_{-0.62}$ &  1.31$^{+1.03}_{-0.62}$ \\
   24.25 &   3.26$^{+1.39}_{-1.01}$ &  3.26$^{+1.39}_{-1.01}$ \\
   24.75 &   6.85$^{+1.85}_{-1.49}$ &  6.85$^{+1.85}_{-1.49}$ \\
   25.25 &   12.73$^{+2.38}_{-2.03}$ &  12.73$^{+2.38}_{-2.03}$ \\
   25.75 &   19.58$^{+2.87}_{-2.52}$ &  19.96$^{+2.93}_{-2.57}$ \\
   26.25 &   31.66$^{+3.55}_{-3.21}$ &  39.24$^{+4.40}_{-3.98}$ \\
\noalign{\smallskip}
\hline
\end{tabular}
\end{center}
\end{table}

\begin{table}
\begin{center}
\caption{Ly$\alpha$ Luminosity Function}
\begin{tabular}{lcccc}
\hline\noalign{\smallskip}
log [L(Ly$\alpha$)] & $\Phi$ (COSMOS)\tablenotemark{a}  & $\Phi$ (ECDF-S)\tablenotemark{a} & $\Phi$ (COSMOS)\tablenotemark{b}  & $\Phi$ (ECDF-S)\tablenotemark{b}  \\
(erg\,s$^{-1}$)  & $(\frac{10^{-4}}{{\rm Mpc^3}(\Delta {\rm log}L)})$ & $(\frac{10^{-4}}{{\rm Mpc^3}(\Delta {\rm log}L)})$ & $(\frac{10^{-4}}{{\rm Mpc^3}(\Delta {\rm log}L)})$ & $(\frac{10^{-4}}{{\rm Mpc^3}(\Delta {\rm log}L)})$  \\
\noalign{\smallskip}\hline
   41.90 &  28.44$^{+6.61}_{-6.28}$  &  33.05$^{+7.28}_{-7.01}$  &  53.49$^{+12.72}_{-12.01}$  &  49.11$^{+10.82}_{-10.41}$   \\
   42.10 &  19.94$^{+4.90}_{-4.59}$  &  19.95$^{+4.87}_{-4.57}$  &  21.51$^{+5.63}_{-5.19}$  &  19.88$^{+5.08}_{-4.71}$ \\
   42.30 &  13.36$^{+3.71}_{-3.36}$  &  12.29$^{+3.49}_{-3.14}$  &  17.91$^{+4.65}_{-4.29}$  &  11.88$^{+3.46}_{-3.09}$ \\
   42.50 &  6.03$^{+2.34}_{-1.92}$  &  6.36$^{+2.38}_{-1.97}$  &  9.23$^{+2.98}_{-2.58}$  &    8.50$^{+2.80}_{-2.41}$ \\
   42.70 &  2.16$^{+1.51}_{-1.01}$  &  1.27$^{+1.26}_{-0.73}$  &  3.02$^{+1.71}_{-1.24}$  &    1.27$^{+1.26}_{-0.73}$  \\
   42.90 &  0.86$^{+1.15}_{-0.58}$  &  1.69$^{+1.37}_{-0.86}$  &  1.72$^{+1.40}_{-0.88}$  &    1.69$^{+1.37}_{-0.86}$  \\
   43.10 &  0.86$^{+1.15}_{-0.58}$  &  0.42$^{+0.98}_{-0.36}$ &  0.86$^{+1.15}_{-0.58}$  &    1.27$^{+1.26}_{-0.73}$  \\
   43.30 &  0.86$^{+1.15}_{-0.58}$  &  ...  &  0.86$^{+1.15}_{-0.58}$  &    ...  \\
   43.50 &  ... & ... &  0.43$^{+0.99}_{-0.36}$  &    ...  \\
\noalign{\smallskip}
\hline
\end{tabular}
\end{center}
\tablenotetext{a}{Aperture-corrected fluxes and completeness curves based on bright stars are used.}
\tablenotetext{b}{Fluxes represented by MAG\_AUTO and completeness curves derived with the reconstructed narrowband image of our LAEs are used.}
\end{table}

\clearpage

\clearpage

\end{document}